\documentclass[prd,email,twocolumn,showpacs,showkeys,preprintnumbers,amsmath,amssymb,nofootinbib]{revtex4-1}

\usepackage{amsmath}
\usepackage{latexsym}

\def\[{\left\lbrack}
\def\]{\right\rbrack}

\def\({\left(}
\def\){\right)}

\def\p{\partial}
\newcommand{\be}{\begin{equation}}
\newcommand{\ee}{\end{equation}}
\newcommand{\ea}{\end{eqnarray}}
\newcommand{\ba}{\begin{eqnarray}}
\newcommand{\vx}{{\vec{x}}}
\newcommand{\vy}{{\vec{y}}}
\newcommand{\dirac}{{\delta(\vx - \vy)}}

\newcommand{\vep}{{\varepsilon}}

\newcommand{\cl}{{\cal L}}
\newcommand{\cg}{{\cal G}}

\begin{document}

\title{Obtaining gauge invariant actions via symplectic embedding formalism}

\author{E. M. C. Abreu$^{a,b,c}$}
\email{evertonabreu@ufrrj.br}
\author{J. Ananias Neto$^c$} 
\email{jorge@fisica.ufjf.br}
\author{A. C. R. Mendes$^c$}
\email{albert@fisica.ufjf.br} 
\author{C. Neves$^d$}
\email{clifford@fat.uerj.br} 
\author{W. Oliveira$^c$}
\email{wilson@fisica.ufjf.br}

\affiliation{${}^{a}$Grupo de F\' isica Te\'orica, Departamento de F\'{\i}sica, Universidade Federal Rural do Rio de Janeiro\\
BR 465-07, 23890-971, Serop\'edica, RJ, Brazil\\
${}^{b}$Centro Brasileiro de Pesquisas F\' isicas (CBPF), Rua Xavier Sigaud 150,\\
Urca, 22290-180, RJ, Brazil\\
${}^{c}$Departamento de F\'{\i}sica, ICE, Universidade Federal de Juiz de Fora,\\
36036-330, Juiz de Fora, MG, Brazil\\
${}^{d}$Departamento de Matem\'atica e Computa\c{c}\~ao, Universidade do Estado do Rio de Janeiro\\
Rodovia Presidente Dutra, km 298, 27537-000, Resende, RJ, Brazil\\
\today\\}
\pacs{11.15.-q; 11.10.Ef; 11.30.Cp}

\keywords{gauge invariance, symplectic embedding, constrained systems}

\begin{abstract}
\noindent The concept of gauge invariance is one of the most subtle and useful concepts in modern theoretical physics.  It is one of the Standard Model cornerstones.  The main benefit due to the gauge invariance is that it can permit the comprehension of difficult systems in physics with an arbitrary choice of a reference frame at every instant of time.  It is the objective of this work to show a path of obtaining gauge invariant theories from non-invariant ones.  Both are named also as first- and second-class theories respectively, obeying Dirac's formalism.  Namely, it is very important to understand why it is always desirable to have a bridge between gauge invariant and non-invariant theories.  Once established, this kind of mapping between first-class (gauge invariant) and second-class systems, in Dirac's formalism can be considered as a sort of equivalence.  This work describe this kind of equivalence obtaining a gauge invariant theory starting with a non-invariant one using the symplectic embedding formalism developed by some of us some years back.  To illustrate the procedure it was analyzed both Abelian and non-Abelian theories.  It was demonstrated that this method is  more convenient than others.  For example, it was shown exactly that this embedding method used here does not require any special modification to handle with non-Abelian systems.  
\end{abstract}

\maketitle

\pagestyle{myheadings}
\markright{{\it Obtaining gauge invariant actions via symplectic embedding formalism}}



\section{Introduction}
\renewcommand{\theequation}{1.\arabic{equation}}
\setcounter{equation}{0}

In this work the symplectic embedding of second-class systems transforming them into gauge theories is discussed \cite{PD}.  Although there is an extensive literature about the subject, we believe that the results, the gauge invariance and the symplectic embedding applications described here are shown in a self-contained and pedestrian way.

The connection between these conversion features can be understood as a kind of physical equivalence.  This connection characterizes both systems as representing the same physical properties \cite{Gil}. The term embedding, used here (and in the literature about the subject) sum up precisely the gist of the procedure.  We will use a general canonical formalism of embedding developed by some of us based on symplectic formalism \cite{ANO}, which embeds a second-class theory 
(the initial action, noninvariant) into one that has gauge invariance. It is important to emphasize here that our embedding leads to a mixed system having both first and second class constraints.   

To perform such conversion task we then used the so-called symplectic embedding formalism developed by some of us in \cite{ANO}.  The main advantage of this method, when compared with other conversion methods, is that it involves the construction of a zero-mode object, which is connected with the symmetries of the final model.  Hence, the embedding method discloses all the hidden symmetries of the original system, since it can be demonstrated that the original and final models present the same equations of motion \cite{Gil}.  The revealing of the set of symmetries of the model is a fundamental issue concerning the fathoming of the original action.  Another advantage is that the method can deal with Abelian and non-Abelian theories.  Namely, nothing needs to be modified or adapted when we use the embedding method to deal with such theories.  In the last two sections we will show that this method can be used by a wider community since we attack fluid systems.

As well known, after the procedure application, where we have first-class constraints, the final system is a gauge invariant theory. The relevance of having a gauge theory, in few words, is that the physically important variables are those that are independent of the local reference frame \cite{ht}.  Whenever a change in the arbitrary reference frame causes a transformation of the variables involved we have the so-called gauge transformation.  This physical variables are then well known as gauge invariant variables.  Such gauge theories and gauge transformations are the cornerstones of the standard model structure, to mention only one of its successful applications.  We will talk with more detail about gauge theories and quantization in a moment.

Since the Hamiltonian formulation is considered by many as the more fundamental formulation of a physical theory, whenever necessary, we also used the Dirac brackets \cite{PD} in order to show exactly the gauge invariance of the final actions obtained.

\bigskip

{\bf Quantization.}  The quantization of gauge theories demands a special care because the presence of  gauge symmetries indicates that exist some superfluous degrees of freedom, which must be eliminated (before or after) with the implementation of a valid quantization process.

The quantization of first-class systems was formulated both in Dirac's  \cite{PD} and path integral  \cite{FADDEEV} points of view. Later on, the path integral analysis was extended by Batalin, Fradkin and Vilkovisky \cite{FRADKIN} in order to preserve the BRST symmetry  \cite{BRST}.

On the other hand, the covariant quantization of second-class systems is, in general, a difficult task because the Poisson brackets are
replaced by Dirac brackets. At the quantum level, the variables become operators and the Dirac brackets become commutators. Hence,
the canonical quantization process is contaminated with serious issues such as ordering operator problems  \cite{order} and anomalies  \cite{RR} in the context of nonlinear constrained systems and chiral gauge theories, respectively. It seems that it is more natural and safe to work out the quantization of second-class systems without invoking Dirac brackets. Actually, it was the strategy followed by many authors over the last decades. The noninvariant system has been embedded in an extended phase space in order to change the second-class nature of constraints into first-class.

In this way, the entire machinery  \cite{BRST,BV} for quantizing first class systems can be used. To implement this concept, Faddeev  \cite{FS} suggests to enlarge the phase space with the introduction of new variables to linearize the system, which were named after them, as the Wess-Zumino (WZ) variables \cite{wz71}. This idea has been embraced by many authors and some methods were proposed and some constraint conversion formalisms, based on the Dirac method  \cite{PD}, were constructed. Among them, the Batalin-Fradkin-Fradkina-Tyutin (BFFT)  \cite{BT,rr100} and the iterative  \cite{IJMP} methods were strong enough to be successfully applied to a great number of important physical systems. Although these techniques share the same conceptual basis  \cite{FS} and follow the Dirac framework  \cite{PD}, these constraint conversion methods were implemented following different directions. Historically, both BFFT and iterative methods were introduced to deal with linear systems such as chiral gauge theories  \cite{IJMP,many3} in order to eliminate the gauge anomaly that hampers the quantization process.

\bigskip

{\bf The Symplectic formalism.}  The technique formulated by Faddeev and Jackiw \cite{FJ} relies on first-order Lagrangians which equations of motion does not generate accelerations.  The brackets involved are called generalized brackets since it can be shown that they coincide with those obtained directly from Dirac's formalism.  So, we can realize that these generalized brackets are linked to the commutators of the quantized theory.  In FJ method the classification of a system as constrained or unconstrained is connected with the behavior of the symplectic two-form.  The outcome of this classification does not necessarily agree with the one given by the Dirac technique.  

\bigskip

{\bf The BFFT technique.} The general canonical quantization formalism due to Batalin, Fradkin, Fradkina and Tyutin (BFFT) \cite{BT} for converting second-class constraints into first-class ones, increases the number of degrees of freedom to include unphysical ones \cite{rr100}.  The obtainment of first-class constraints is carried out in an iterative process \cite{BN}.  To summarize we can say that the first correction of the constraints is linear in the auxiliary variables.  The second correction is quadratic and so on.  For systems with just linear constraints we have to perform the conversion to first-class ones.  For nonlinear constraints, we have more than one iteration and the BFFT formalism shows its problems since the first step of the process does not precisely fix the solution that will be used in the next steps.  Some solutions of this problem were published but they cannot be applied for all cases.


\bigskip

{\bf The Noether dualization method.}  Recently, the so-called gauging iterative Noether dualization 
method  \cite{ainrw} has been shown to thrive in establishing some dualities between models  \cite{iw}.
This method hinges on the traditional concept of a local lifting of a global symmetry and it may be
realized by an iterative embedding of Noether counterterms.  However, this
method provides a strong suggestion of duality since it has been shown to give
the expected result in the paradigmatic duality between the so-called self-dual model  \cite{tpn} and the Maxwell-Chern-Simons
theory in three dimensions duality.  This correspondence was first established by Deser and Jackiw  \cite{dj}
and using the parent action approach  \cite{suecos}.

\bigskip

{\bf The paper.}   In section 2, we review the symplectic embedding formalism in order to settle the
notation and to make the reader to familiarize with the fundamentals of the formalism.
In section 3, we will initiate with the Proca model in order to set up the general ideas discussed in Section 2.  We will see that a different zero-mode can bring us an equivalent action different from the literature.
After that, we apply this
formalism in two important models in high energy physics. The first one, in section 4, is the nonlinear sigma model (NLSM) \cite{NLSM}, which is an important
theoretical laboratory to learn the basics about asymptotically free field theories, as dynamical mass generation, confinement,
and topological excitations, which is expected in the realistic world of four-dimensional non-Abelian gauge theories. The second, in section 5, is the
bosonized chiral Schwinger model (CSM), which has attracted too much attention over the last decade, mainly in the context of string
theories \cite{STRING}, and also due to the huge progress in the understanding of the physical meaning of anomalies in quantum field theories
achieved through the intensively study of this model. The gauge invariant version of the model was obtained. 
Section 6 is devoted to an application of the formalism to the non-Abelian Proca model.
We will show that this gauge-invariant formalism does not require modifications to deal with non-Abelian models as demanded by the BFFT method.  In section 7, we enlarged the method boundaries performing the embedding of a fluid dynamical model, showing that the method is not restricted only to the theoretical ones.
In section 8, we analyze the symmetries of the rotational fluid model with a new extra term, which introduced a dissipative force into the system.  The objective is to promote an approximation to reality where we always have dissipation. Whenever convenient, the gauge symmetry will be investigated from the Dirac point of view.
Our concluding observations and final comments are given in Section 9.

\section{The  symplectic embedding formalism}\label{s2}
\renewcommand{\theequation}{2.\arabic{equation}}
\setcounter{equation}{0}

The formalism is developed based on the symplectic framework  \cite{FJ,BC}, that is a modern way to handle with constrained systems. The basic object behind this formalism is the presymplectic matrix, i.e., if this matrix is singular, the model presents a symmetry, if it is not singular, the Dirac brackets can be obtained. In this way, we have  proposed to turn nonsingular presymplectic matrix to a singular one. This procedure will be carried out through the introduction of both the arbitrary functions that depend on the original coordinates and the WZ variables into the first-order Lagrangian. To appreciate this point, a brief review of the symplectic formalism will be provided in this section.   After that, general ideas of the symplectic gauge-invariant formalism will be explained. This formalism, differently from the BFFT and other iterative constraint conversion methods, does not require special modifications of the formalism to convert Abelian or non-Abelian sets of second-class constraints into first-class ones.  We think that this is the main advantage of this formalism.

In the following lines, as we said above, we will try to keep this paper self-contained.  We will follow closely the ideas contained in
 \cite{amnot}.

\bigskip

Let us now consider a general noninvariant mechanical model whose dynamics is governed by a Lagrangian
${\cal L}(a_i,\dot a_i,t)$, (with $i=1,2,\dots,N$), where $a_i$ and $\dot a_i$
are the space and velocity variables, respectively.   Following the symplectic method the zeroth-iterative
first-order one-form Lagrangian is written as
 \begin{equation}
\label{2000}
{\cal L}^{(0)}dt = A^{(0)}_\theta d\xi^{(0)\theta} - V^{(0)}(\xi)dt\,\,,
\end{equation}
with arbitrary one form $A^{(0)} = A^{(0)}_{\theta}\,d\xi^{(0)\,\theta}$ and the presymplectic variables are defined as
\be
\xi^{(0)\theta} =  \left\{ \begin{array}{ll}
                               a_i, & \mbox{with $\theta=1,2,\dots,N $} \\
                               p_i, & \mbox{with $\theta=N + 1,N + 2,\dots,2N ,$}
                           \end{array}
                     \right.
\ee
where $A^{(0)}_\theta$ are the canonical momenta and $V^{(0)}$ is defined as being the presymplectic potential. From the Euler-Lagrange equations of motion we have that $$f^{(0)}_{\theta\beta}\,\dot{\xi}^{(0)\beta}\,=\,\frac{\partial V}{\partial \xi^{(0)\theta}}$$ whose solutions rely on the invertibility of $f^{(0)}_{\theta\beta}$.  The problem is that it is an impossible task if there are constraints involved.  They would provide a singular matrix since 
\be
\label{AAAA}
\dot{\xi}^{(0)\beta}\,=\,(f^{(0)}_{\theta\beta})^{-1}\frac{\partial V}{\partial \xi^{(0)\theta}}\,\,.  
\ee
It is easy to see that the Hamiltonian corresponding to the Lagrangian (\ref{2000}) is $V^{(0)}$.  So, Eqs. (\ref{AAAA}) are also Hamiltonian-type
$$\dot{\xi}^{(0)\beta}=\{V^{(0)},\xi^{(0)\beta}\}=\frac{\partial V}{\partial \xi^{(0)\theta}}\,\{\xi^{(0)\theta},\xi^{(0)\beta}\}.$$  But the generalized bracket is defined to be $\{\xi^{(0)\theta},\xi^{(0)\beta}\}=(f^{(0)}_{\theta\beta})^{-1}$.  The equations of motion (\ref{AAAA}) can be written in Hamiltonian form, with the presymplectic potential $V^{(0)}$ playing the role of Hamilton function.

The presymplectic tensor is obtained as
\begin{eqnarray}
\label{2010}
f^{(0)}_{\theta\beta} = {\partial A^{(0)}_\beta\over \partial \xi^{(0)\theta}}
-{\partial A^{(0)}_\theta\over \partial \xi^{(0)\beta}}\,\,.
\end{eqnarray}
If the two-form
$$f \equiv \frac{1}{2}f_{\theta\beta}d\xi^\theta \wedge d\xi^\beta$$

\noindent is singular, the presymplectic matrix (\ref{2010}) has a zero-mode $(\nu^{(0)})$ that generates a new constraint when multiplied by (contracted with) the gradient of the presymplectic potential,
\begin{equation}
\label{2020}
\Omega^{(0)} = \nu^{(0)\theta}\frac{\partial V^{(0)}}{\partial\xi^{(0)\theta}}\,\,.
\end{equation}
This constraint will be introduced into the zeroth-iterative one-form Lagrangian equation (\ref{2000}) through a Lagrange multiplier $\eta$, generating the next one
\begin{eqnarray}
\label{2030}
{\cal L}^{(1)}dt &=& A^{(0)}_\theta d\xi^{(0)\theta} + d\eta\Omega^{(0)}- V^{(0)}(\xi)dt,\nonumber\\
&=& A^{(1)}_\gamma d\xi^{(1)\gamma} - V^{(1)}(\xi)dt,\end{eqnarray}
with $\gamma=1,2,\dots,(2N + 1)$ and
\begin{eqnarray}
\label{2040}
V^{(1)}&=&V^{(0)}|_{\Omega^{(0)}= 0},\nonumber\\
\xi^{(1)_\gamma} &=& (\xi^{(0)\theta},\eta),\\
A^{(1)}_\gamma &=&(A^{(0)}_\theta, \Omega^{(0)})\,\,.\nonumber
\end{eqnarray}
As a consequence, the first-iterative presymplectic tensor can be computed as
\begin{eqnarray}
\label{2050}
f^{(1)}_{\gamma\beta} = {\partial A^{(1)}_\beta\over \partial \xi^{(1)\gamma}}
-{\partial A^{(1)}_\gamma\over \partial \xi^{(1)\beta}} \,\,.
\end{eqnarray}
If this tensor is nonsingular, the iterative process stops and the Dirac brackets
 among the phase space variables are obtained from the inverse matrix
 $(f^{(1)}_{\gamma\beta})^{-1}$ and consequently, the Hamiltonian equation of
 motion can be formulated and solved, as discussed in  \cite{gotay}. It is well known
 that a physical system can be described at least classically in terms of a presymplectic
 manifold ${\cal M}$. From a physical point of view, ${\cal M}$ is the phase space of the system while
 a nondegenerate closed 2-form $f$ can be identified as being the Poisson bracket. The
 dynamics of the system is  determined just specifying a real-valued function (Hamiltonian)
$H$ on the phase space, {\it i.e.}, one of these real-valued function
solves the Hamiltonian equation, namely,
\be \label{2050a1}
\iota(X)f=dH, \ee
and the classical dynamical trajectories of the
system in the phase space are obtained. It is important to mention
that if $f$ is nondegenerate, the equation (\ref{2050a1}) has a very unique
solution. The nondegeneracy of $f$ means that the linear map
$\flat:TM\rightarrow T^*M$ defined by $\flat(X):=\flat(X)f$ is an
isomorphism.  So, equation (\ref{2050a1}) can be solved uniquely
for any Hamiltonian $(X=\flat^{-1}(dH))$. On the other hand, the
tensor has a zero-mode and a new constraint arises, indicating
that the iterative process goes on until the presymplectic matrix
becomes nonsingular or singular. If this matrix is nonsingular,
the Dirac brackets will be determined naturally. 
In \cite{gotay}, the authors consider in detail the case when $f$ is degenerated. 

The main idea of this embedding formalism is to introduce extra fields into the model in order to obstruct the solutions of the Hamiltonian equations of motion.
We introduce two arbitrary functions that hinge on the original phase space and on the WZ variables, namely, $\Psi(a_i,p_i)$ and $G(a_i,p_i,\eta)$, into the first-order one-form Lagrangian as follows
\be
\label{2060a}
{\tilde{\cal L}}^{(0)}dt = A^{(0)}_\theta d\xi^{(0)\theta} + \Psi d\eta - {\tilde V}^{(0)}(\xi)dt,
\ee
with
\be
\label{2060b}
{\tilde V}^{(0)} = V^{(0)} + G(a_i,p_i,\eta),
\ee
where the arbitrary function $G(a_i,p_i,\eta)$ is expressed as an expansion in terms of the WZ field, given by
\begin{equation}
\label{2060}
G(a_i,p_i,\eta)=\sum_{n=1}^\infty{\cal G}^{(n)}(a_i,p_i,\eta),,
\end{equation}
where ${\cal G}^{(n)}(a_i,p_i,\eta)\sim\eta^n \,,$
and satisfies the following boundary condition
\begin{eqnarray}
\label{2070}
G(a_i,p_i,\eta=0) = 0.
\end{eqnarray}
The presymplectic variables were extended until they also encompass the WZ variable $\tilde\xi^{(0)\tilde\theta} = (\xi^{(0)\theta},\eta)$ (with ${\tilde\theta}=1,2,\dots,2N+1$) and the first-iterative presymplectic potential becomes
\begin{equation}
\label{2075}
{\tilde V}^{(0)}(a_i,p_i,\eta) = V^{(0)}(a_i,p_i) + \sum_{n=1}^\infty{\cal G}^{(n)}(a_i,p_i,\eta).
\end{equation}
In this context, the new canonical momenta are
\be
{\tilde A}_{\tilde\theta}^{(0)} = \left\{\begin{array}{ll}
                                  A_{\theta}^{(0)}, & \mbox{with $\tilde\theta$ =1,2,\dots,2N}\\
                                  \Psi, & \mbox{with ${\tilde\theta}$= 2N + 1}
                                    \end{array}
                                  \right.
\ee
and the new presymplectic tensor is given by
\begin{equation}
{\tilde f}_{\tilde\theta\tilde\beta}^{(0)} = \frac {\partial {\tilde A}_{\tilde\beta}^{(0)}}{\partial \tilde\xi^{(0)\tilde\theta}} - \frac {\partial {\tilde A}_{\tilde\theta}^{(0)}}{\partial \tilde\xi^{(0)\tilde\beta}},
\end{equation}
that is
\be
\label{2076b}
{\tilde f}_{\tilde\theta\tilde\beta}^{(0)} =
\begin{pmatrix}
 { f}_{\theta\beta}^{(0)} & { f}_{\theta\eta}^{(0)}
\cr { f}_{\eta\beta}^{(0)} & 0
\end{pmatrix}.
\ee

To sum up, basically, we have two steps: the first one is addressed to compute $\Psi(a_i,p_i)$ while the second one is dedicated to the calculation of $G(a_i,p_i,\eta)$. In order to begin with the first step, we impose that this new presymplectic tensor (${\tilde f}^{(0)}$) has a zero-mode $\tilde\nu$, consequently, we obtain the following condition
\begin{equation}
\label{2076}
\tilde\nu^{(0)\tilde\theta}{\tilde f}^{(0)}_{\tilde\theta\tilde\beta} = 0\,\,.
\end{equation}
At this point, $f$ becomes degenerated and in consequence, we introduced an obstruction to solve the Hamiltonian equation of motion given by equation (\ref{2050a1}). Assuming that the zero-mode $\tilde\nu^{(0)\tilde\theta}$ is
\begin{equation}
\label{2076a}
\tilde\nu^{(0)}=
\begin{pmatrix}
\mu^\theta & 1
\end{pmatrix},
\end{equation}
and using the relation given in (\ref{2076}) combined with (\ref{2076b}), we have a system of equations,
\be
\label{2076c}
\mu^\theta{ f}_{\theta\beta}^{(0)} + { f}_{\eta\beta}^{(0)} = 0,
\ee
where
\be
{ f}_{\eta\beta}^{(0)} =  \frac {\partial A_\beta^{(0)}}{\partial \eta} - \frac {\partial \Psi}{\partial \xi^{(0)\beta}}\,\,.
\ee
The matrix elements $\mu^\theta$ are chosen in order to disclose the desired gauge symmetry. Note that in this formalism the zero-mode $\tilde\nu^{(0)\tilde\theta}$ is the gauge symmetry generator. It is worth to mention that this feature is important because it opens up the possibility to disclose the desired hidden gauge symmetry from the noninvariant model.  From relation (\ref{2076}) some differential equations involving $\Psi(a_i,p_i)$ are obtained, {\it i. e.}, the equation (\ref{2076c}), and after a straightforward computation, $\Psi(a_i,p_i)$ can be determined.

In order to compute $G(a_i,p_i,\eta)$ following the second step of the method, it is mandatory that no constraints arise from the contraction of the zero-mode $(\tilde\nu^{(0)\tilde\theta})$ with the gradient of the potential ${\tilde V}^{(0)}(a_i,p_i,\eta)$. This condition generates a general system of differential equations, which is
\begin{eqnarray}
\label{2080}
&&\tilde\nu^{(0)\tilde\theta}\frac{\partial {\tilde V}^{(0)}(a_i,p_i,\eta)}{\partial{\tilde\xi}^{(0)\tilde\theta}}\,=\, 0,\\
&&\mu^\theta \frac{\partial {V}^{(0)}(a_i,p_i)}{\partial{\xi}^{(0)\theta}} + \mu^\theta \frac{\partial {\cal G}^{(1)}(a_i,p_i,\eta)}{\partial{\xi}^{(0)\theta}} \nonumber \\
\,&+&\, \mu^\theta\frac{\partial {\cal G}^{(2)}(a_i,p_i,\eta)}{\partial{\xi}^{(0)\theta}} + \dots
\,+\,\frac{\partial {\cal G}^{(1)}(a_i,p_i,\eta)}{\partial\eta} \nonumber \\
&+& \frac{\partial {\cal G}^{(2)}(a_i,p_i,\eta)}{\partial\eta} + \dots = 0\;\; 
\mbox{}
\end{eqnarray}
and that allows us to compute all the correction terms ${\cal G}^{(n)}(a_i,p_i,\eta)$ as functions of $\eta$. Notice that this polynomial expansion in terms of $\eta$ is equal to zero.  Consequently, all the coefficients for each order in $\eta$ must be identically zero.
Given this, each correction term as function of $\eta$ is determined. For a linear correction term, we have
\begin{equation}
\label{2090}
\mu^\theta\frac{\partial V^{(0)}(a_i,p_i)}{\partial\xi^{(0)\theta}} + \frac{\partial{\cal
 G}^{(1)}(a_i,p_i,\eta)}{\partial\eta} = 0\,\,.
\end{equation}
For a quadratic correction term, we have
\begin{equation}
\label{2095}
{\mu}^{\theta}\frac{\partial{\cal G}^{(1)}(a_i,p_i,\eta)}{\partial{\xi}^{(0)\theta}} + \frac{\partial{\cal G}^{(2)}(a_i,p_i,\eta)}{\partial\eta} = 0.
\end{equation}
From these both equations, a recursive equation for $n\geq 2$ can be chosen as,
\begin{equation}
\label{2100}
{\mu}^{\theta}\frac{\partial {\cal G}^{(n - 1)}(a_i,p_i,\eta)}{\partial{\xi}^{(0)\theta}} + \frac{\partial{\cal
 G}^{(n)}(a_i,p_i,\eta)}{\partial\eta} = 0,
\end{equation}
which permits us to compute the remaining correction terms as functions of $\eta$. This iterative process is successively repeated until (\ref{2080}) becomes identically zero.  Consequently, the extra term $G(a_i,p_i,\eta)$ is obtained explicitly. Then, the gauge invariant Hamiltonian, identified as being the presymplectic potential, is obtained from
\begin{equation}
\label{2110}
{\tilde{\cal  H}}(a_i,p_i,\eta) = V^{(0)}(a_i,p_i) + G(a_i,p_i,\eta),
\end{equation}
and the zero-mode ${\tilde\nu}^{(0)\tilde\theta}$ is identified as the generator of an infinitesimal gauge transformation, given by
\begin{equation}
\label{2120}
\delta{\tilde\xi}^{\tilde\theta} = \varepsilon{\tilde\nu}^{(0)\tilde\theta},
\end{equation}
where $\varepsilon$ is an infinitesimal parameter.

In the following sections, we will apply the symplectic gauge-invariant formalism in some second-class constrained Hamiltonian systems, Abelian and non-Abelian.

\section{The Abelian Proca model}
\renewcommand{\theequation}{3.\arabic{equation}}
\setcounter{equation}{0}

Let us start with a simple Abelian
case which is the Proca model whose dynamics is ruled by the Lagrangian density,
\begin{equation}
\label{Proca1}
{\cal L} = - \,\,\frac{1}{4}F_{\mu\nu}F^{\mu\nu} + \frac{1}{2}\,m^2\,A^{\mu}A_{\mu},
\end{equation}
where $m$ is the mass, $g_{\mu\nu} = diag(+---)$ and (from now on) $F_{\mu\nu} = \partial_{\mu}A_{\nu} - \partial_{\nu}A_{\mu}$.
As well known, the mass term breaks the gauge invariance of the usual Maxwell's theory. Hence, the Lagrangian density above represents a
second-class system.

To begin with the symplectic embedding procedure the Lagrangian density is reduced to its first-order form as
\begin{equation}
\label{Proca3}
{\cal L}^{(0)} = \pi^{i}\dot{A_{i}} - V^{(0)},
\end{equation}
where the presymplectic potential is
\be
\label{Proca3a}
V^{(0)} = \frac{1}{2}{\pi_{i}}^2 + \frac 14 F_{ij}^2 + \frac{1}{2}\,m^2\,{A_{i}}^2 - A_{0}(\partial_i\pi^{i} + \frac{1}{2}\,m^2\,A_{0}) ,
\ee
where $\pi_i=\dot A_i - \partial_iA_0$.  From now on $\partial_i=\frac{\partial}{\partial x^i}$ and the dot denote space and time derivatives, respectively. The presymplectic coordinates are $\xi_\alpha^{(0)}=(A_i,\pi_i,A_0)$ and the corresponding one-form canonical momenta are given by $a_{A_i}^{(0)} \,=\, \pi_i$ and $a_{\pi^i}^{(0)} \,=\, a_{A_0}^{(0)} = 0$.
The zeroth-iteration presymplectic matrix is
\begin{equation}
f^{(0)} = \left(
\begin{array}{ccc}
0           & -\delta_{ij} & 0 \\
\delta_{ji}&         0     & 0 \\
0           &         0     & 0
\end{array}
\right)\,\delta^{(3)}({\vec x} - {\vec y}),
\end{equation}
which is a singular matrix. It has a zero-mode that generates the constraint $\Omega = \partial_i\pi^i + m^2A_0$,
identified as the Gauss law.  We will include this constraint into the canonical part of the first-order Lagrangian ${\cal L}^{(0)}$
in (\ref{Proca3}) introducing a Lagrangian multiplier ($\beta$).  The first-iterated Lagrangian can be written in terms of $\xi_\alpha^{(1)} = (A_i,\pi_i,A_0, \beta)$ as
${\cal L}^{(1)} = \pi^{i}\dot{A_{i}} + \Omega\dot{\beta} - V^{(1)}$,
with the following presymplectic potential,
\be
\label{Proca6a}
V^{(1)} = \frac{1}{2}{\pi_{i}}^2 + \frac 14 F_{ij}^2 + \frac{1}{2}\,m^2\,\({A_{0}}^2 + {A_{i}}^2\) - A_0\Omega.
\ee
The first-iterated presymplectic matrix, computed as
\begin{equation}
f^{(1)}=\left(
\begin{array}{cccc}
0           & -\delta_{ij}  &  0   &   0 \\
\delta_{ji} &         0     &   0    &    \partial^y_i \\
0         &         0     &   0    &     m^2   \\
0       &        -\partial^x_j    &  -m^2     &     0
\end{array}
\right)\,\delta^{(3)}({\vec x} - {\vec y}),
\end{equation}
is a nonsingular matrix.  Consequently, the Proca model is not a gauge invariant field theory. The Poisson brackets among the phase space fields can be obtained from the inverse of the presymplectic matrix.
The Hamiltonian,
\ba
\label{hamil01}
{\cal H} = V^{(1)}|_{\Omega=0} &=& \frac 12 \pi_i^2 - \frac {1}{2 m^2}\pi_i\partial^i\partial_j\pi^j + \frac 14 F_{ij}^2 + \frac 12 m^2 A_i^2,\nonumber\\
&=& \frac 12 \pi_iM^i_j\pi^j + \frac 14 F_{ij}^2 + \frac 12 m^2 A_i^2,
\ea
where the phase space metric is $M^i_j = g^i_j - \frac{\partial^i\partial_j}{m^2}$,
which completes the noninvariant analysis.

The goal here is to disclose the gauge symmetry hidden inside the model. Both arbitrary functions, $\Psi$ and $G$ rely on both the original phase space variables and the WZ variable $(\theta)$.  The former ($\Psi$) is introduced into the kinetic sector and the later (G), within the potential sector of the first-order Lagrangian. The process starts with the computation of $\Psi$ and it ends up the calculation of $G$.

To clarify, the problem is that we have a second-class theory that is not (obviously) gauge-invariant.  So, we have to make something if we want to transform this theory into a gauge invariant one.  Thus, we have to introduce ``something" new so that, interacting with the other fields and/or with itself, promotes a gauge-invariance ``process" inside the theory.  This ``something" is the $\theta$-field and the ``process" is the symplectic embedding fornalism.  However, it is important to notice that this $\theta$-field, although now it is part of the symplectic coordinates of the extended phase space, it does not have a conjugated canonical momentum.  Consequently, we believe that, from the outset, this $\theta$-field can not be interpreted as the St\"uckelberg field.  Even if we consider that the canonical momentum conjugated to the St\"uckelberg field is usually canceled by the Hamilton's equation of motion, the final result is different from the one obtained by the symplectic embedding.  Note that the structures of the first-order Lagrangians for both methods are different since in the St\"uckelberg method \cite{banban} we have the pair $\pi_{\theta} \dot{\theta}$ and in the symplectic embedding we have $\Psi \dot{\theta}$, where $\Psi$ is the arbitrary function introduced above.

The first-order Lagrangian ${\cal L}^{(0)},$ given in Eq. (\ref{Proca3}), with the arbitrary terms, given by,
\begin{equation}
\label{Proca6b}
{\tilde{\cal L}}^{(0)} = \pi^{i}\dot{A_{i}} + \dot\theta\,(\Psi\,+\,\gamma) - {\tilde V}^{(0)},
\end{equation}
where
\ba
\label{Proca6c}
{\tilde V}^{(0)} &=&  \frac{1}{2}{\pi_{i}}^2 + \frac 14 F_{ij}^2  + \frac{1}{2}\,m^2\,{A_{i}}^2 - A_{0}(\partial_i\pi^{i} + \frac{1}{2}\,m^2\,A_{0}) \nonumber \\ 
&+& G\,+\,\frac 12\,k\,\gamma\gamma,
\ea
and $\Psi\equiv\Psi(A_i,\pi_i,A_0,\theta)$ and $G\equiv G(A_i,\pi_i,A_0,\theta)$ are the arbitrary functions to be determined as well as the constant $k$. Now, the presymplectic fields are ${\tilde\xi}^{(0)}_\alpha=(A_i,\pi_i,A_0,\theta,\gamma)$ while the presymplectic matrix is
\begin{widetext}
\be
\label{matrix00}
\tilde{f}^{(0)} =
\begin{pmatrix}
0 & 0 & 0 & \frac{\partial\Psi_y}{\partial A^x_0} & 0 \cr
0 & 0 & - g_{ij}\delta^{(3)}(\vec x - \vec y) & \frac{\partial\Psi_y}{\partial A^x_i} & 0  \cr 
0 & g_{ji}\delta^{(3)}(\vec x - \vec y) & 0 & \frac{\partial\Psi_y}{\partial \pi^x_i} & 0 \cr 
- \frac{\partial\Psi_x}{\partial A^y_j} & - \frac{\partial\Psi_x}{\partial \pi^y_j} & - \frac{\partial\Psi_x}{\partial A^y_0} & f_{\theta_x\theta_y} & - \delta^{(3)}(\vec x - \vec y) \cr
0 & 0 & 0 & \delta^{(3)}(\vec x - \vec y) & 0
\end{pmatrix}
\ee
\end{widetext}
with
\be
\label{matrix01}
f_{\theta_x\theta_y} = \frac{\partial \Psi_y}{\partial \theta_x} - \frac{\partial \Psi_x}{\partial \theta_y},
\ee
where $\theta_x \equiv \theta(x)$, $\theta_y \equiv \theta(y)$, $\Psi_x \equiv \Psi(x)$ and $\Psi_y \equiv \Psi(y)$.
Note that $\tilde f$ is a $9 \times 9$ matrix with two space indexes in each entry. There is also an implicit time dependence, which comes from the coordinates and momenta. In the representation of $\tilde f$, described above, some zeros are actually zero columns, zero lines or zero matrices.

The corresponding zero-mode $\nu^{(0)}(\vec x)$, the generator of the symmetry, satisfies the following relation,
\be
\label{matrix02}
\int \,\, d^3y \,\,\nu^{(0)}_\alpha(\vec y)\,\,f_{\alpha\beta}(\vec x - \vec y)= 0\,\,.
\ee
The zero-mode does not generate a new constraint. However, it determines the arbitrary function $\Psi$ and consequently, it obtains the gauge invariant reformulation of the model. We will scrutinize the gauge symmetry related to the following zero-mode,
\be
	\tilde \nu  = \begin{pmatrix} 1 & 0_{1 \times 3} & 0_{1 \times 3} & 0 & b \end{pmatrix} 
\ee
to be the zero-mode of $\tilde{f}^{(0)}$, where $b$ is a constant. 

The constraint generated by $\tilde{\nu}$ is $\Omega = -\partial_i\pi^i - m^2 A_0$. As we will see, $\tilde \nu$ will produce a constraint which is equal to $\Omega$ when $\gamma = \theta = 0$.

Following the procedure we know that $\tilde \nu$ is a zero-mode of $\tilde f$, one condition for $\Psi$ is found, which is
\be
	\frac {\delta \Psi(\vy)}{\delta A^0(\vx)} = -b \dirac.
	\label{psi_a0proca}
\ee

The constraint appears from the contraction,
\ba
	\tilde \Omega (\vx) &=& \int d^3y \; \tilde \nu^\alpha(\vx) \frac{\delta \tilde V(\vy)}{\delta \tilde \xi^\alpha (\vx)} \\
&=& -\partial_i \pi^i - m^2A_0 + \int d^3y \; \frac{\delta G(\vy)}{\delta A_0 (\vx)}  + bk \gamma, \nonumber 
	\label{modconstproca}
\ea
or, for short, $\tilde \Omega = \Omega + G_0 + bk\gamma$, where $G_0$ is implicitly defined.

Now we add $\dot \lambda \tilde \Omega$ to $\tilde \cl$ and consider $\lambda$ as a new independent field, that is, a Lagrange multiplier. Hence,
\be
	\tilde \cl^{(1)} = \pi^i \dot A_i + (\Psi + \gamma) \dot \theta + \dot \lambda \tilde \Omega - \tilde V.
\ee
The presence of the constraint inside the kinetic part of the Lagrangian allows us to remove it from the potential part. Nevertheless, this common procedure will not help us here.  Therefore no change happened in the potential.

Setting $\tilde \xi^{(1)\alpha} = (A^0, A^i, \pi^i, \theta, \gamma, \lambda)$ as the new symplectic coordinates, where from now on $\alpha = 1,2,...,10$, and with the help of equation (\ref{psi_a0proca}), the symplectic matrix is
\begin{widetext}
\be
\tilde f^{(1)} = 
\begin{pmatrix}
0 & 0 & 0 & -b\delta^{(3)} & 0 & \frac {\delta G_0(\vy)}{\delta A_0(\vx)} - m^2\delta^{(3)} \\[0.1in] \cr
0 & 0 & -g_{ji} \delta^{(3)} & \frac{\delta \Psi(\vy)}{\delta A^i(\vx)} & 0 & \frac {\delta G_0(\vy)}{\delta A^i(\vx)} \\[0.1in] \cr
0 & g_{ij} \delta^{(3)} & 0 & \frac{\delta \Psi(\vy)}{\delta \pi^i(\vx)} & 0 & \frac {\delta G_0(\vy)}{\delta \pi^i(\vx)} - {\partial}_i^y \delta^{(3)} \\[0.1in] \cr
b\delta^{(3)} & - \frac{\delta \Psi(\vx)}{\delta A^j(\vy)} & - \frac{\delta \Psi(\vx)}{\delta \pi^j(\vy)} & \Theta_{xy} & - \delta^{(3)} & \frac {\delta G_0(\vy)}{\delta \theta (\vx)} \\[0.1in] \cr
0 & 0 & 0 & \delta^{(3)} & 0 & bk \delta^{(3)} \\[0.1in] \cr
- \frac {\delta G_0(\vx)}{\delta A_0(\vy)} + m^2 \delta^{(3)} & \frac {\delta G_0(\vx)}{\delta A^j(\vy)} & {\partial}_j^x \delta^{(3)} - \frac {\delta G_0(\vx)}{\delta \pi^j(\vy)} & - \frac {\delta G_0(\vx)}{\delta \theta(\vy)} & -bk \delta^{(3)} & 0
\end{pmatrix}.
\ee
\end{widetext}
For the sake of clarity, it is convenient to use the notation $\delta^{(3)}$ instead of $\dirac$.

We can select two independent zero-modes to become the infinitesimal gauge generators, which are
\ba
	\tilde \nu_{(\theta)} & = & \begin{pmatrix} a_0 & a \partial^i & c \partial^i & -kb & 0 & 1  \end{pmatrix}, \nonumber \\
	\tilde \nu_{(\gamma)} & = & \begin{pmatrix} 1 & 0_{1 \times 3} & 0_{1 \times 3} & 0\;\; & b\; & 0 \end{pmatrix} = \begin{pmatrix}
 \tilde \nu & 0 \end{pmatrix}.
	\label{zmproca}
\ea
The values of the constants $a_0, \; a$ and $c$ can be freely chosen.  We have to remember that different choices directly correspond to different gauge generators. As it will be shown, the value of $b$ is also free. 

Naturally, other zero-mode structures are possible, some of which entail correspondence to both Wess-Zumino fields in each set.

For $\tilde \nu_{(\gamma)}$ just one condition is necessary to assure its zero-mode feature, namely,
\be
	\frac{\delta G_0 (\vy)}{\delta A_0 (\vx)} = (m^2-b^2k)\dirac.
	\label{G0proca}
\ee

This zero-mode is a generator of gauge symmetries.  Therefore no new constraint may arise from its contraction with the gradient of the potential. By equations (\ref{modconstproca}) and (\ref{zmproca}), we see that this condition is automatically fulfilled.

There is a set of nontrivial equations that needs to be satisfied in order to $\tilde \nu_{(\theta)}$ be a zero-mode of $\tilde f^{(1)}$. Instead of evaluating them now, it seems to be easier to demand that $\tilde \nu_{(\theta)}$ may not give rise to a new constraint. Hence,
\ba
	0 & = & \int d^3y \; \tilde \nu^\alpha_{(\theta)} (\vx) \; \frac{\delta \tilde V (\vy)}{\delta \tilde \xi^{(1) \alpha}(\vx)} \nonumber \\
	\label{Gproca}
	& = & \int d^3y \left \{ a_0 \dirac ( -\partial_i \pi^i - m^2A_0 ) \right. \nonumber \\ 
&+& \left. a\partial^i_x \dirac ( \partial^j F_{ij} - m^2 A_i )       \right. \\
	&+& \left. \; c\partial^i_x  \dirac ( -\pi_i + \partial_i A_0 ) + \rho^\mu_x \frac{\delta G(\vy)}{\delta A^\mu(\vx)} \right. \nonumber \\
&+& \left. c\partial^i_x \frac{\delta G(\vy)}{\delta \pi^i(\vx)} - kb \frac{\delta G(\vy)}{\delta \theta(\vx)} \right \}. \nonumber
\ea
The index $x$ on $\partial^i$ means that the derivative must be evaluated with respect to $x$ (i.e., $\partial_x^i \equiv \partial / \partial x_i$), and 
\be
	\rho^\mu_x \equiv ( a_0, a\partial^i_x)
\ee

Equation (\ref{Gproca}) can be solved by considering $G$ as a power series of $\theta$ (and its spatial derivatives). Let $\cg_n$ be proportional to $\theta^n$, so $G = \sum_n \cg_n$. The condition $G(\theta = 0) = 0$ leads to $n \ge 1$. Hence,
\be
	\cg_1 = \frac{\theta}{kb} ( -a_0 \partial_i\pi^i -m^2\rho^\mu A_\mu - c\partial^i\pi_i + c\partial^i\partial_i A_0).
\ee

The terms $\frac{\delta G(\vy)}{\delta A^\mu(\vx)}$ and $\frac{\delta G(\vy)}{\delta \pi^i(\vx)}$ do not contribute to the computation of $\cg_1$.  But they do contribute to others $\cg_n$'s. After some straightforward calculations, one can find $\cg_2$ (without surface terms) as
\be
	\cg_2 = - \frac 1{2(kb)^2} \{ c(2 a_0 + c) \partial_i \theta \partial^i \theta  + m^2 \rho^\mu \theta \rho_\mu \theta \}.
\ee 

The absence of $A^\mu$ and $\pi^i$ in $\cg_2$ implies that $\cg_n = 0$ for all $n \ge 3$. Thus the function $G$ is completely known and we can write down the expression for $G_0$, which is,
\be
	G_0(\vx) \equiv \int d^3y \; \frac {\delta G(\vy)}{\delta A^0 (\vx)} = \frac 1 {kb} (c\partial^i \partial_i \theta - m^2 a_0 \theta).
	\label{G0proca2}
\ee
Substituting this result in equation (\ref{G0proca}), we have that
$k = \frac {m^2}{b^2}$.
This fixes $k$ as dependent of $b$.

Our next and final step in order to prescribe the gauge embedded Lagrangian is to find $\Psi$. This can be done by demanding that $\tilde \nu_{(\theta)}$ be a zero-mode of $\tilde f^{(1)}$. Using (\ref{G0proca2}) and (3.20) we have that, 
\be
	c\partial_j^x \dirac + \frac{m^2}{b} \frac{\delta \Psi(\vx)}{\delta A^j(\vy)} = 0,
	\label{psiaproca}
\ee
\be
	- a\partial_j^x \dirac + \frac {m^2}b \frac {\delta \Psi (\vx)}{\delta \pi^j(\vy)} + \partial_j^x \dirac = 0,
	\label{psipiproca}
\ee
\ba
&-& b a_0 \dirac + a\partial_x^i \frac{\delta \Psi(\vy)}{\delta A^i(\vx)} + c \partial^i_x \frac {\delta \Psi (\vy)}{\delta \pi^i(\vx)} - \frac {m^2}b \Theta_{xy} \nonumber \\
&-& \frac{\delta G_0 (\vx)}{\delta \theta (\vy)} = 0\,\,.
	\label{psithetaproca}
\ea
	With equations (\ref{psi_a0proca}) and (\ref{psiaproca}-\ref{psithetaproca}), up to an additive function just of $\theta$ (action surface term), $\Psi$ can be determined as,
\be
	\Psi = - \frac b {m^2} \{ m^2 A_0 + c \partial_i A^i + (1 - a)\partial^i \pi_i \}.
\ee

Note that we can withdraw the term $\dot \lambda \tilde \Omega$ from $\tilde \cl^{(1)}$ without changing the dynamics. One can always redo the symplectic algorithm and find again the constraint $\tilde \Omega$.   This will lead us back to $\tilde \cl$. By varying $\tilde \cl$ with respect to $\pi_i$ and using Euler-Lagrange equations we find
\be
	\pi_i = \partial_i A_0 - \dot A_i + \frac b {m^2} \{(1-a) \partial_i \dot \theta + (a_0 + c) \partial_i \theta \}.
\ee

Note that the momenta are not the original ones (which are $F_{i0}$), but when $\theta$ is eliminated they are recovered.

Also from the Euler-Lagrange equations, we have that 
$\gamma = \frac {b^2}{m^2} \dot \theta$.
Thus, eliminating $\pi_i$ and $\gamma$, the Lagrangian $\tilde \cl$ can be expressed by
\ba
	\tilde \cl & = & - \frac 14 F_{\mu \nu} F^{\mu \nu} + \frac {m^2}2 A^\mu A_\mu + \frac b {m^2} \left \{ -m^2 A_0 \dot \theta \right. \nonumber \\
&+& \left. (1-a) \partial_i \dot \theta (\partial^i A_0 - \dot A^i) + a_0 \theta (\partial^i \partial_i A_0 - \partial_i \dot A^i ) \right. \nonumber \\
&+& \left. \theta m^2 \rho^\mu A_\mu \right \} \\[0.1in]
	\label{procaembedL}
&+& \frac {b^2}{m^4} \left \{ \frac 32 (1-a)^2 \partial_i \dot \theta \partial^i \dot \theta - \frac 12 a_0^2 \partial_i \theta \partial^i \theta \right. \nonumber \\
&+& \left. (1-a) (a_0 + c) \partial_i \dot \theta \partial^i \theta + \frac{m^2}2 \rho^\mu \theta \rho_\mu \theta \right \} + \frac {b^2}{2m^2}\dot \theta \dot \theta. \nonumber
\ea

From the components of $\tilde \nu_\theta$ and $\tilde \nu_\gamma$ the infinitesimal gauge generators of the theory are obtained as 
\ba
	\delta_\vep A_0 & = & \vep a_0 -  \dot \vep, \nonumber \\
	\delta_\vep A^i & = & - a \partial^i \vep,  \\
	\delta_\vep \theta & = & - \frac {m^2}b \vep. \nonumber
\ea
where $\varepsilon$ is an infinitesimal time-dependent parameter. 

The symplectic formalism assures us that $\tilde \cl$ is invariant under the above transformations for any constants $b$, $a_0$ and $a$ (assuming they have proper dimensions, which are squared mass, mass and unit respectively). 

Usually, terms with more than two derivatives in the Lagrangian are not welcomed, these can be avoided by fixing $a = 1$.

If one wants an explicit Lorentz invariance, the constants need to be fixed as $b= m^2$, $a = 1$ and $a_0 = 0$ (alternatively, $b$ could also be $-m^2$). With these values, the Lagrangian turn out to have a St\"uckelberg aspect, that is
\be
	\tilde \cl  =  - \frac 14 F_{\mu \nu} F^{\mu \nu} + \frac {m^2}2 A^\mu A_\mu - m^2A^\mu \partial_\mu \theta + \frac {m^2}2 \partial^\mu \theta \partial_\mu \theta.
\ee


The Lagrangian in Eq. (3.32) is not the most general one that can be achieved with the symplectic embedding method. Others structures of the zero-modes $\tilde \nu_{(\theta)}$ and $\tilde \nu_{(\gamma)}$ are also possible, and their components, together with the components of $\tilde \nu$, could also be field dependent.

If we analyze this specific example (Abelian Proca model), we can see that the BFFT embedding in the Hamiltonian approach is the analogue of the St\"uckelberg mechanism in the Lagrangian version.  
The extra field in the BFFT version is just the St\"uckelberg field.  
This was discussed in \cite{banban}.  
This shows clearly what we said above about the fact that our $\theta$-field is not the St\"uckelberg field.

Following the symplectic embedding formalism, the zero-mode $\tilde \nu^{(0)}$ is the generator of the infinitesimal gauge transformation $(\delta{\cal O}=\varepsilon\tilde\nu^{(0)})$, given by,
\begin{eqnarray}
\label{Proca16}
\delta A_{\mu} = - \partial_{\mu}\varepsilon,\qquad
\delta\theta = -\,\varepsilon\,\,.\nonumber 
\end{eqnarray}
Indeed, for the above transformations the invariant Hamiltonian,
identified as being the presymplectic potential ${\tilde V}^{(0)}$, changes as $\delta{\cal H} = 0$.  It is a very easy task to show the invariance of (3.35) under the above gauge transformations.   Consequently, there is no need to carry out the Dirac analysis.

\section{The $O(N)$ invariant nonlinear sigma model}
\renewcommand{\theequation}{4.\arabic{equation}}
\setcounter{equation}{0}

The $O(N)$ nonlinear sigma model (NLSM) in two dimensions is a free field theory for the multiplet $\sigma_a\equiv
(\sigma_1,\sigma_2,\dots,\sigma_n)$ satisfying a nonlinear constraint $\sigma_a^2=1$. This model has its dynamics governed by the
Lagrangian density
\begin{equation}
{\cal L}=\frac{1}{2}\,\partial_\mu\sigma^a\partial^\mu\sigma_a
- \frac{1}{2}\,\lambda\,\bigl(\sigma^a\sigma_a - 1\bigr),
\label{3001}
\end{equation}
where $\mu=0,1$ and $a$ is an index related to the $O(N)$ symmetry group.

The original second order Lagrangian in the velocity, given in (\ref{3001}), is reduced into
a first-order form, given by,
\begin{equation}
\label{3002}
{\cal L}^{(0)} = \pi_a\dot{\sigma}^a - V^{0},
\end{equation}
with $V^{(0)} = \frac{1}{2}\,\pi^2_a + \frac{1}{2}\,\lambda\,\bigl(\sigma^2_a - 1\bigr) - \frac{1}{2}\,{\sigma^\prime_a} ^2$.  The presymplectic coordinates are
$\xi_\alpha^{(0)}=(\sigma_a,\pi_a,\lambda)$. The presymplectic tensor given by Eq. (\ref{2010}) is 
\begin{equation}
\label{3003}
f^{(0)} = \left(
\begin{array}{ccc}
0           & -\delta_{ab} & 0 \\
\delta_{ba} &         0     & 0 \\
0           &         0     & 0
\end{array}
\right)\,\delta(x-y).
\end{equation}
Since this matrix is singular, it has a zero-mode, 
$\nu^{(0)} = \left(
\begin{array}{ccc}
{\bf 0} & {\bf 0} & 1
\end{array}
\right)$.
Contracting this zero-mode with the gradient of the presymplectic potential $V^{(0)}$, given above, the constraint  obtained is,
$\Omega_1 = \sigma^2_a- 1$.
The first-iteration Lagrangian is
\be
\label{3005}
{\cal L}^{(1)} = \pi_a\dot{\sigma}^a + \Omega_1\dot{\rho} - V^{(1)}\mid_{\Omega_1=0},
\ee
with $V^{1}\mid_{\Omega_1=0} =\frac{1}{2}\,\pi^2_a  + \frac{1}{2}\,{\sigma^\prime}^2_a$.
The presymplectic coordinates are $\xi_\alpha^{(1)}=(\sigma_a,\pi_a,\rho)$ with the following one-form canonical momenta,
$A_{\sigma_a}^{(1)} = \pi_a$, 
$A_{\pi_a}^{(1)} = 0$ and $A_{\rho}^{(1)} = \bigl(\sigma^2_a - 1\bigr)$.

The corresponding presymplectic tensor $f^{(1)}$ given by,
\begin{equation}
\label{3008}
f^{(1)}=\left(
\begin{array}{ccc}
0           & -\delta_{ab} & \sigma_a \\
\delta_{ab} &         0     & 0 \\
-\sigma_b           &         0     & 0
\end{array}
\right)\,\delta(x-y),
\end{equation}
is singular.   Consequently, it has a zero-mode that generates a new constraint, $\Omega_2 = \sigma_a\pi^a$ and the
second-iteration Lagrangian is obtained as
\begin{equation}
\label{3010}
{\cal L}^{(2)} = \pi_a \dot{\sigma}^a + \dot{\rho}\bigl(\sigma^2_a - 1\bigr)+ \dot{\zeta}(\sigma_a\pi^a) - V^{(2)},
\end{equation}
with $V^{(2)}$ = $V^{(1)}\mid_{\Omega_1=0}$. The enlarged presymplectic coordinates are $\xi_\alpha^{(2)}=(\sigma_a,\pi_a,\rho,\zeta)$ and the
new one-form canonical momenta are
\begin{eqnarray}
\label{formula22}
A_{\sigma_a}^{(2)} = \pi_a, \quad
A_{\pi_a}^{(2)} = 0, \quad
A_{\rho}^{(2)} = \sigma^2_a - 1,\quad
A_{\zeta}^{(2)} = \sigma_a \pi^a. \nonumber
\end{eqnarray}
The corresponding matrix $f^{(2)}$ is
\begin{equation}
\label{3011}
f^{(2)}=\left(
\begin{array}{cccc}
0           & -\delta_{ab}  &  \sigma_a   &   \pi_a \\
\delta_{ba} &         0     &   0    &    \sigma_a \\
-\sigma_b         &         0     &   0    &     0   \\
-\pi_b       &        -\sigma_b    &   0    &     0
\end{array}
\right)\,\delta(x-y),
\end{equation}
which is a nonsingular matrix. 
This means that the NLSM is not a gauge invariant theory.  Now, the original phase
space will be extended with the introduction of a WZ field.  Let us introduce $\Psi(\sigma_a,\pi_a,\theta)$ and  $G(\sigma_a,\pi_a,\theta)$, into the
first-order Lagrangian as follows,
\be
\label{3012a}
{\tilde {\cal L}}^{(0)} = \pi_a\dot{\sigma}^a + \Psi\dot{\theta} - {\tilde V}^{(0)},
\ee
where the presymplectic potential is
\be
\label{3013a}
{\tilde V}^{(0)} = \frac{1}{2}\,\pi^2_a + \frac{1}{2}\,\lambda\,\bigl(\sigma^2_a - 1\bigr) + \frac{1}{2}\,{\sigma^\prime_a} ^2 + G(\sigma_a,\pi_a,\theta),
\ee
with $G(\sigma_a,\pi_a,\theta)$ satisfying the relations given in Eqs. (\ref{2060}) and (\ref{2070}).

The presymplectic coordinates are $\tilde \xi_\alpha^{(0)}=(\sigma_a,\pi_a,\lambda,\theta)$
with the following one-form canonical momenta,
\begin{eqnarray}
\label{3015}
\tilde A_{\sigma_a}^{(0)} = \pi_a, \quad
\tilde A_{\pi_a}^{(0)} = 0, \quad
\tilde A_{\lambda}^{(0)} = \frac 12 (\sigma^2_a - 1), \quad
\tilde A_{\theta}^{(0)} = 0. \nonumber \\
\end{eqnarray}

The corresponding matrix $\tilde f^{(0)}$, given by
\begin{equation}
\label{3016}
\tilde f^{(0)} =
\begin{pmatrix}
0  & -\delta_{ab}  &  0  &   \frac{\partial\Psi_y}{\partial\sigma^{x}_a}
\cr \delta_{ba} &  0 & 0  & \frac{\partial\Psi_y}{\partial\pi^{x}_a}
\cr 0 & 0 & 0  & \frac{\partial\Psi_y}{\partial\lambda^{x}}
\cr - \frac{\partial\Psi_x}{\partial\sigma^{y}_b} & - \frac{\partial\Psi_x}{\partial\pi^{y}_b}  & -\frac{\partial\Psi_x}{\partial\lambda^{y}} & f_{\theta_x\theta_y}
\end{pmatrix}
\delta(x-y),
\end{equation}

\noindent must be singular.   This settles down the dependence relations between arbitrary function $\Psi$, namely,
$\frac{\partial\Psi_y}{\partial\lambda^{x}_a}=0$, so, $\Psi\equiv\Psi(\sigma_a,\pi_a,\theta)$.  
This matrix has a zero-mode.

Considering the symmetry generated by the following zero-mode,
$\nu^{(0)}=\left(
\begin{array}{cccc}
{\bf 0} & \sigma_a & 0 & 1
\end{array}
\right)$.
Since this zero-mode and the presymplectic matrix (\ref{3016}) satisfy the relation (\ref{matrix02}), $\Psi$ is determined as $\Psi = {1\over2}\,\sigma_a^2+c$, where $c$ is a constant parameter. 

We know that no more constraints are generated by the contraction of the zero-mode with the gradient of the potential.  The correction terms as functions of $\theta$ can be explicitly computed. The first-order correction term in $\theta$, ${\cal G}^{(1)}$, determined after an integration process, is
${\cal G}^{(1)}(\sigma_a,\pi_a,\theta) = - \sigma_a\pi_a\theta$.
Substituting this expression into Eq. (\ref{3013a}), the new Lagrangian is
\begin{equation}
\label{3019}
\tilde {\cal L}^{(0)} = \pi_a\dot{\sigma}^a + \Psi\dot{\theta} - \frac{1}{2}\,{\sigma^\prime}^2_a - \frac{1}{2}\,\pi^2_a - \frac{1}{2}\,\lambda\,\bigl(\sigma^2_a - 1\bigr)  + \sigma_a\pi_a\theta.
\end{equation}

However, the model is not yet gauge invariant because the contraction of the zero-mode $\nu^{(0)}$ with the gradient of the potential
$V^{0}$ produces a non zero value, indicating that it is necessary to compute the remaining correction terms ${\cal G}^{(n)}$ as functions of
$\theta$. It can be carried out just demanding that the zero-mode does not generate a new constraint. It allows us to determine the second order
correction term given by ${\cal G}^{(2)} = + \frac {1}{2} \sigma_a^2 \theta^2$.
Substituting this result into the first-order Lagrangian (\ref{3019}), we have that,
\ba
\label{3021}
\tilde {\cal L}^{(0)} &=& \pi_a\dot{\sigma}^a + \Psi\dot{\theta} - \frac{1}{2}\,{\sigma^\prime} ^2_a -
\frac{1}{2}\,\lambda\,\bigl(\sigma^2_a - 1\bigr)  - \frac{1}{2}\,\pi^2_a  \nonumber \\
&+& \sigma_a\pi^a\theta - \frac {1}{2} \sigma_a^2 \theta^2.
\ea
Now the zero-mode $\nu^{(0)}$ does not produce new constraints.   Consequently, the model has a symmetry and all correction terms ${\cal G}^{(n)}$ with
$n\geq 3$ are zero.

We can recover the invariant second order Lagrangian from its first-order form.
To this end, the canonical momenta must be eliminated from the Lagrangian (\ref{3021}).   The canonical momenta are computed as $\pi_a = \dot \sigma_a + \sigma_a\theta$.
Inserting this result into the first-order gauge-invariant Lagrangian (\ref{3021}), we have the second order Lagrangian as
\begin{equation}
\label{3023}
\tilde {\cal L} = \frac{1}{2}\,\partial_\mu\sigma_a\partial^\mu\sigma^a - (\dot{\sigma_a}\sigma^a)\theta -
\frac{1}{2}\bigl(\sigma^2_a - 1\bigr)\lambda ,
\end{equation}
with the following gauge invariant Hamiltonian,
\begin{equation}
\label{3024}
\tilde {\cal H} = \frac{1}{2}\,\pi^2_a  + \frac{1}{2}\,{\sigma^\prime} ^2_a - (\sigma_a\pi^a)\theta +
\frac{1}{2}\lambda \bigl(\sigma^2_a - 1\bigr)+ \frac{1}{2}\,\sigma^2_a\theta^2 .
\end{equation}

\noindent From the Dirac point of view, $\Omega_1$ arises as a secondary constraint from the temporal stability imposed on the primary constraints, $\pi_\lambda$ and
$\pi_\theta$,  and plays the role of the Gauss law, which generates the time independent gauge transformation. 

To proceed the quantization, we recognize the states of physical interest as those that are annihilated by $\Omega_1$. 

The infinitesimal gauge transformations $\delta\tilde \xi_\alpha^{(0)}=\varepsilon \nu^{(0)}$, are 
\begin{eqnarray}
\delta \sigma_a &=& 0, \qquad \qquad \delta \pi_a = \varepsilon \sigma_a\,\,, \nonumber \\ 
\delta \lambda &=& 0, \qquad \qquad\:\: \delta \theta = \varepsilon\,\,.  \nonumber
\end{eqnarray} 
Concerning these transformations the Hamiltonian changes as $\delta{\cal H} = 0$.

Similar results were also obtained in the literature using different methods based on Dirac's constraint
framework \cite{WN,BGB,JW1,JW2,HKP,NW}. 

To disclose the hidden symmetry of the NLSM lying on the original phase space $(\sigma_a,\pi_a)$, 
we can use the Dirac method to obtain the set of constraints of the gauge invariant NLSM.  It was described by the Lagrangian (\ref{3023}) and Hamiltonian (\ref{3024}), given by, $\phi_1 = \pi_\lambda$, $\phi_2 = - \frac 12(\sigma^2_a - 1)$,
and $\varphi_1 = \pi_\theta$ and $\varphi_2 = \sigma_a\pi_a - \sigma_a^2\theta$, where $\pi_\lambda$ and $\pi_\theta$ are the canonical momenta conjugated to $\lambda$ and $\theta$, respectively. The corresponding Dirac matrix is singular.  However, there are nonvanishing Poisson brackets among some constraints, indicating that there are both second-class and first-class constraints.  This problem is solved separating the second-class constraints from the first-class ones through constraint analysis. The set of first-class constraints is
\begin{eqnarray}
\label{3028}
\chi_1 = \pi_\lambda, \qquad
\chi_2 = - \frac 12 (\sigma^2_a - 1) + \pi_\theta \,\,,
\end{eqnarray}
while the set of second-class constraints is
$\varsigma_1 = \pi_\theta$ and $\varsigma_2 = \sigma_a\pi_a - \sigma^2_a\theta$.
Since the second-class constraints are assumed to be equal to zero in a strong way \cite{NM} the Dirac
brackets are constructed as
\begin{eqnarray}
\label{3030}
\lbrace \sigma_i(x), \sigma_j(y)\rbrace^* &=& 0, \nonumber \\
\lbrace \sigma_i(x), \pi_j(y)\rbrace^* &=& \delta_{ij}\,\delta(x-y)\,\,, \\
\lbrace \pi_i(x), \pi_j(y)\rbrace^* &=& 0.\nonumber
\end{eqnarray}
Hence, the gauge invariant Hamiltonian is rewritten as
\begin{eqnarray}
\label{3031}
\tilde {\cal H} &=& \frac{1}{2}\pi^2_a + \frac{1}{2}\,{\sigma^\prime} ^2_a
- \frac{1}{2}\frac{(\sigma_a\pi^a)^2}{\sigma_a\sigma^a} + \frac{\lambda}{2}(\sigma^2_a -1)\nonumber\\
&=& {1\over 2} \pi_iM_{ij}\pi_j + \frac{1}{2}\,{\sigma^\prime}^2_a + \frac{\lambda}{2}(\sigma_a^2 - 1),
\end{eqnarray}
where the phase space metric $M_{ij}$, given by $M_{ij} = \delta_{ij} - \frac {\sigma_i\sigma_j}{\sigma_k^2}$,
which is a singular matrix.   The set of first-class constraints becomes
\ba
\label{3033}
\chi_1 = \pi_\lambda\,\,, \qquad
\chi_2 = - \frac 12(\sigma^2_a - 1).
\ea
Note that the constraint $\chi_2$, originally a second-class constraint, becomes the generator of gauge symmetries, satisfying the first-class property
$\{\chi_2, \tilde{H} \} = 0$.
Due to this result, the infinitesimal gauge transformations are computed as
\ba
\label{3036}
\delta \sigma_a &=& \varepsilon\lbrace \sigma_a,\chi_2\rbrace = 0,\nonumber\\
\delta \pi_a &=& \varepsilon\lbrace\pi_a,\chi_2\rbrace=\varepsilon \sigma_a,\\
\delta\lambda &=& 0.\nonumber
\ea
where $\varepsilon$ is an infinitesimal parameter. It is easy to verify that the Hamiltonian (\ref{3031}) is invariant under these
transformations because $\sigma_a$ are eigenvectors of the phase space metric ($M_{ij}$) with null eigenvalues. In this section we
reproduced the results originally obtained in  \cite{KR} using an alternative point of view.

\section{The gauge invariant bosonized Chiral Schwinger Model (CSM)}
\renewcommand{\theequation}{5.\arabic{equation}}
\setcounter{equation}{0}

It has been shown over the last decade that anomalous gauge theories in two dimensions can be consistently and unitarily quantized for
both Abelian \cite{RR,JR,many} and non-Abelian \cite{RR1,LR} cases. In this scenario, the two dimensional model that has been extensively
studied is the CSM. We start with the following Lagrangian density of the bosonized CSM with $a > 1$,
\ba
{\cal L} &=& -\frac 14 \, F_{\mu\nu}\,F^{\mu\nu} +
            \frac{1}{2}\,\partial_\mu\phi\,\partial^{\mu}\phi +
            q\,\left(g^{\mu\nu} -\,\epsilon^{\mu\nu}\right)\,
            \partial_{\mu}\phi\,A_{\nu} \nonumber \\
            &+&
           \frac{1}{2}\,q^2a\,A_{\mu}A^{\mu}\,.
\label{00000}
\ea

\noindent Now, we have that $g_{\mu\nu} =
\mbox{diag}(+1,-1)$ and $\epsilon^{01} = -\epsilon^{10} = \epsilon_{10} = 1$.   
The Lagrangian in (\ref{00000}), can be reduced into its first-order as follows,
\be
\label{mitra1}
{\cal L}^{(0)} =\pi _\phi \dot{\phi} + \pi _1 \dot{A_1} - U^{(0)},
\end{equation}
where the zeroth-iterative presymplectic potential $U^{(0)}$ is
\begin{eqnarray}
\label{mitra2}
U^{(0)}&=& {1\over 2}(\pi _1^2 +\pi _\phi ^2 +\phi ^{\prime 2}) - A_0\big[ \pi _1^\prime +
{1\over2}q^2(a-1)A_0  \nonumber \\
&+& q^2A_1 + q\pi _\phi + q\phi^\prime \big]\nonumber \\
&-& A_1 \big[-q\pi_\phi -{1\over 2}q^2(a+1)A_1 - q\phi^\prime \big]\,\,.
\end{eqnarray}
The zeroth-iterative presymplectic variables are
$\xi _\alpha^{(0)}=( \phi ,\pi _\phi,A_0 ,A_1 ,\pi _1)$ with
the following one-form canonical momenta $A_\alpha $,
\begin{eqnarray}
\label{00050}
A_\phi ^{(0)} = \pi _\phi, \qquad
A_{A_1}^{(0)} = \pi_1, \qquad
A_{\pi _\phi}^{(0)} = A_{A_0}^{(0)}=A_{\pi _1}^{(0)} = 0. \nonumber \\ 
\end{eqnarray}
The zeroth-iterative presymplectic tensor can be obtained as
\begin{equation}
\label{00060}
f^{(0)}(x,y)= \left( \begin{array}{ccccc}
0 & -1 & 0 & 0 & 0 \\
1 & 0 & 0 & 0 & 0 \\
0 & 0 & 0 & 0 & 0 \\
0 & 0 & 0 & 0 & -1 \\
0 & 0 & 0 & 1 & 0
\end{array} \right )\delta(x - y).
\end{equation}
This matrix is obviously singular.  Thus, it has a zero-mode that generates a constraint when contracted with the gradient of the potential $U^{(0)}$, given by,
\begin{eqnarray}
\label{00080}
\Omega _1 &=& \nu_\alpha ^{(0)}{{\partial U^{(0)}}\over {\partial \xi _\alpha ^{(0)}}}\nonumber \\
&=&\pi _1^\prime +q^2(a-1)A_0 + q^2A_1 + q\pi _\phi + q\phi ^\prime,
\end{eqnarray}
that is identified as the Gauss law, which satisfies the Poisson algebra, $\lbrace\Omega _1(x),\Omega _1(y)\rbrace = 0$.  The first-iterative Lagrangian $L^{(1)}$ is
\begin{equation}
\label{00090}
{\cal L}^{(1)} =\pi _\phi \dot{\phi} + \pi _1 \dot{A_1} + \Omega_1 \dot {\eta } - U^{(1)},
\end{equation}
with the first-order presymplectic potential given by
\begin{eqnarray}
\label{00100}
U^{(1)}&=& {1\over 2}(\pi _1^2 +\pi _\phi ^2 +\phi ^{\prime 2}) \nonumber \\
&-& A_0 \big[ \pi _1^\prime + {1\over2}q^2(a-1)A_0  + q^2A_1 + q\pi _\phi + q\phi^\prime \big]\nonumber \\
&-& A_1 \big[-q\pi_\phi -{1\over 2}q^2(a+1)A_1 - q\phi^\prime \big],
\end{eqnarray}
where $U^{(1)}=U^{(0)}$. Therefore, the presymplectic variables become $\xi _\alpha^{(1)}=( \phi ,\pi _\phi, A_0 ,A_1 ,\pi _1, \eta )$
with the following one-form canonical momenta,
\begin{eqnarray}
\label{00110}
A_\phi ^{(1)} &=& \pi _\phi ,\qquad  A_{A_1}^{(1)} = \pi _1 ,\nonumber \\
A_{A_0}^{(1)} &=& A_{\pi _\phi}^{(1)}= A_{\pi _1}^{(1)}= 0, \\
A_{\eta}^{(1)}&=& \pi _1^\prime +q^2(a-1)A_0 + q^2A_1 + q\pi _\phi + q\phi ^\prime. \nonumber
\end{eqnarray}
The corresponding matrix $f^{(1)}$ is then
\begin{equation}
\label{00120}
f^{(1)}(x, y)= \left ( \begin{array}{cccccc}
0 & -1 & 0 & 0 & 0 & q\partial_y \\
1 & 0 & 0 & 0 & 0 & q \\
0 & 0 & 0 & 0 & 0 & Q \\
0 & 0 & 0 & 0 & -1 & q^2 \\
0 & 0 & 0 & 1 & 0 & \partial_y \\
-q\partial _x & - q & -Q & -q^2 & - \partial_x & 0
\end{array} \right )\delta (x-y),
\end{equation}
where $Q=q^2(a-1)$ and this is a nonsingular matrix. 
This means that the model is not a gauge invariant theory.

We have now two arbitrary functions $\Psi(\phi,\pi_\phi,A_0,A_1,\pi_1,\theta)$ and $G(\phi,\pi_\phi,A_0,A_1,\pi_1,\theta)$,
depending on both the original phase space
variables and the WZ variable $\theta$.  The first-order Lagrangian can be rewritten as
\begin{equation}
\label{00300}
{\tilde {\cal L}}^{(0)} =\pi _\phi \dot{\phi} + \pi _1 \dot{A_1} + \dot\theta\Psi - {\tilde U}^{(0)},
\end{equation}
where
\begin{eqnarray}
\label{00310}
&&{\tilde U}^{(0)}\,=\, {1\over 2}(\pi _1^2 +\pi _\phi ^2 +\phi ^{\prime 2}) \nonumber \\
&-&  A_0\big[ \pi _1^\prime + {1\over2}q^2(a-1)A_0 + q^2A_1 + q\pi _\phi + q\phi^\prime \big]\nonumber \\
&-& A_1 \big[-q\pi_\phi -{1\over 2}q^2(a+1)A_1 - q\phi^\prime \big]  \\
&+& G(\phi , \pi _\phi, A_0, A_1, \pi _1, \theta )\,\,. \nonumber
\end{eqnarray}

The enlarged presymplectic variables are now ${\tilde \xi} _\alpha^{(0)}=( \phi ,\pi _\phi, A_0 ,A_1 ,\pi _1, \theta )$ with the following one-form canonical momenta
\begin{eqnarray}
\label{00320}
{\tilde A}_\phi ^{(0)} &=& \pi _\phi ,\qquad
{\tilde A}_{A_1}^{(0)} = \pi _1 , \qquad
{\tilde A}_{A_0}^{(0)} = {\tilde A}_{\pi _\phi}^{(0)}={\tilde A}_{\pi _1}^{(0)} = 0, \nonumber \\
&\mbox{and}& \quad {\tilde A}_{\theta}^{(0)} = \Psi\,\,. 
\end{eqnarray}
The corresponding presymplectic matrix ${\tilde f}^{(0)}$ reads
\ba
\label{00325}
{\tilde f}^{(0)}&=&
\begin{pmatrix}
0 & -1 & 0 & 0 & 0 & \frac{\partial\Psi^y}{\partial \phi^x}
\cr 1 & 0 & 0 & 0 & 0 & \frac{\partial\Psi^y}{\partial \pi^x_\phi}
\cr 0 & 0 & 0 & 0 & 0 & \frac{\partial\Psi^y}{\partial A^x_0}
\cr 0 & 0 & 0 & 0 & -1 & \frac{\partial\Psi^y}{\partial A^x_1}
\cr 0 & 0 & 0 & 1 & 0 & \frac{\partial\Psi^y}{\partial \pi^x_1}
\cr - \frac{\partial\Psi^x}{\partial \phi^y} & - \frac{\partial\Psi^x}{\partial \pi^y_\phi}& -\frac{\partial\Psi^x}{\partial A^y_0} &
- \frac{\partial\Psi^x}{\partial A^y_1} & -
\frac{\partial\Psi^x}{\partial \pi^y_1}  & f_{\theta_x\theta_y}
\end{pmatrix} \nonumber \\
&\cdot& \delta (x-y),  
\ea

\noindent where
\be
\label{00325a}
f_{\theta_x\theta_y} = \frac{\partial \Psi_y}{\partial \theta_x} - \frac{\partial \Psi_x}{\partial \theta_y},
\ee
with $\theta_x \equiv \theta(x)$, $\theta_y \equiv \theta(y)$, $\Psi_x \equiv \Psi(x)$ and $\Psi_y \equiv \Psi(y)$. Note that this matrix is singular since $\frac{\partial\Psi^x}{\partial A^y_0} = 0$. 
So, $\Psi\equiv\Psi(\phi,\pi_\phi,A_1,\pi_1,\theta)$.

We will now investigate the symmetry connected to the following zero-mode, 
$\bar\nu^{(0)} =
\begin{pmatrix}
q & - q\partial_x & 1 & \partial_x & - q^2  & - 1
\end{pmatrix}$,
with bar representing a transpose matrix.

Now we have to multiply the zero-mode above by the presymplectic matrix (\ref{00325}). Hence,
some equations arise and after an integration $\Psi$ is determined as $\Psi = \pi^\prime_1 + q\phi^\prime + q\pi_\phi + q^2A_1$.  The presymplectic matrix (\ref{00325}) is rewritten as
\begin{equation}
\label{00330}
{\tilde f}^{(0)}= \left ( \begin{array}{cccccc}
0 & -1 & 0 & 0 & 0 &  q\partial_y\\
1 & 0 & 0 & 0 & 0 & q\\
0 & 0 & 0 & 0 & 0 & 0\\
0 & 0 & 0 & 0 & -1 & q^2 \\
0 & 0 & 0 & 1 & 0 & \partial _y\\
- q\partial_x & - q & 0 & - q^2 & - \partial_x & 0
\end{array} \right )\delta(x-y)
\end{equation}

\noindent which is obviously singular and it has a zero-mode that, by construction, is given by the one above.

The first-order correction term in $\theta $, ${\cal G}^{(1)}$, is determined by,
\ba
\label{00350}
{\cal G}^{(1)}(\phi , \pi_\phi, A_1, \pi_1, A_0, \theta ) = - \Omega_1 \theta + q^2(a-1)A^{\prime}_1\theta -
q^2\theta \pi _1, \nonumber \\
\ea
after an integration. Substituting this expression into the equation (\ref{00300}), the new Lagrangian is
\begin{equation}
\label{00360}
{\tilde {\cal L}}^{(0)} = \pi _\phi \dot{\phi} + \pi _1 \dot{A_1} + \Psi\dot {\theta } - {\tilde U}^{(0)},
\end{equation}
with
\begin{eqnarray}
\label{00361}
&&{\tilde U}^{(0)} \,=\, {1\over 2}(\pi _1^2 +\pi _\phi ^2 +\phi ^{\prime 2}) \nonumber \\
&-& A_0\big[ \pi _1^\prime +{1\over2}q^2(a-1)A_0 + q^2A_1 + q\pi _\phi + q\phi^\prime \big]\nonumber \\
&-& A_1 \big[-q\pi_\phi -{1\over 2}q^2(a+1)A_1 - q\phi^\prime \big] \\
&-&\Omega_1 \theta + q^2(a-1)\theta^\prime A_1 - q^2\theta \pi _1\,\,. \nonumber
\end{eqnarray}

The Lagrangian in (\ref{00360}) is not yet gauge invariant because the zero-mode $\bar\nu^{(0)}$ still generates new constraints, given by
\begin{equation}
\label{00370}
\nu_\alpha^{(1)}{{\partial {\tilde U}^{(0)}}\over {\partial \xi _\alpha ^{(0)}}} = q^2(a-1)\theta^{\prime\prime} -
q^2(a-1) \theta + q^4\theta\,\,.
\end{equation}

\noindent  The second order correction term ${\cal G}^{(2)}$ is,
\begin{eqnarray}
\label{00380}
& &\nu_\alpha^{(0)}{{\partial {\tilde U}^{(0)}}\over {\partial \xi _\alpha ^{(0)}}} \\
&=& - q^2(a-1)\theta + q^2(a-1)\theta^{\prime\prime} +
q^4\theta - {{\partial {\cal G}^{(2)}}\over {\partial \theta }}= 0, \nonumber \\
& &\Rightarrow\,{\cal G}^{(2)} = - {1\over 2}\;\; q^2(a-1){\theta ^{\prime}}^2 + {1\over 2}q^4{\theta }^2 - {1\over 2}q^2(a-1)\theta ^2. \nonumber 
\mbox{}
\end{eqnarray}

Hence, the first-order Lagrangian (\ref{00360}) becomes
\begin{equation}
\label{00390}
{\tilde {\cal L}}^{(0)} = \pi _\phi \dot{\phi} + \pi _1 \dot{A_1} + \Psi\dot\theta - {\tilde U}^{(0)},
\end{equation}
with the new presymplectic potential
\begin{eqnarray}
\label{00391}
&&{\tilde U}^{(0)} \,=\, {1\over 2}(\pi _1^2 +\pi _\phi ^2 +\phi ^{\prime 2}) \nonumber \\
&-& A_0\big[ \pi _1^\prime +{1\over2}q^2(a-1)A_0 + q^2A_1 + q\pi _\phi + q\phi^\prime \big]\nonumber \\
&-& A_1 \big[-q\pi_\phi - {1\over 2}q^2(a+1)A_1 - q\phi^\prime \big] \\
&-& \Omega_1\theta + q^2(a-1)\theta A^\prime_1 - q^2\theta \pi _1 \nonumber \\
&-&{1\over 2}\;\; q^2(a-1){\theta ^{\prime}}^2 + {1\over 2}q^4{\theta }^2 - {1\over 2}q^2(a-1)\theta ^2\,\,. \nonumber
\end{eqnarray}

\noindent and the respective second-order Lagrangian is
\ba
\label{secondV}
{\cal L}&=& \frac 12 \(\dot{\phi}^2 - {\phi'}^2 \) + \frac 12 \theta^{'2} + \theta' \dot{\theta}' + \dot{\phi}\,\dot{\theta} \nonumber \\
&+&\({A'}_0 - \dot{A}_1\) \[\frac 12 \({A'}_0 - \dot{A}_1\) + \theta' + \dot{\theta}' \] \nonumber \\
&+& q \[\phi' \dot{\theta} + \( \phi' - \dot{\phi} \) \theta + \(\dot{\phi} + \phi'\) \( A_0 - A_1 \)\] \nonumber \\
&+& q^2 \[ a\({\theta'}^2 + \theta^2 \) + \(\theta + \theta' \) \dot{\theta} - \theta' \theta - \frac 12 {\theta'}^2 \right. \\
&+& \left. \(\dot{A}_1 - 3 A_1 \) \theta + A_0 \(\dot{\theta} - {\theta'} \) + a A_0 \theta + \frac 12 \( A_0^2 + A_1^2 \) \] \nonumber \\
&+& \frac 12 q^2 (a-1) \[ A^3_0 - A_1^2 - \theta {A'}_1 \] \,\,.\nonumber
\ea

\noindent The contraction of the zero-mode $\bar\nu^{(0)}$ with the new presymplectic potential above does not produce a new constraint. So, all correction
terms ${\cal G}^{(n)}$ with $n \geq 3$ are zero. The infinitesimal gauge transformations originated from the zero-mode $(\delta\xi_i=\varepsilon \nu^{(0)})$ are
\ba
\label{00395}
\delta \phi &=& q\varepsilon, \qquad \qquad \quad \!\!\!
\delta \pi_\phi =  q \varepsilon^{\prime},\nonumber \\
\delta A_0 &=& \varepsilon, \qquad \qquad \quad
\delta A_1 = - \varepsilon^{\prime},\nonumber \\
\delta \pi_1 &=& - q^2\varepsilon, \qquad \qquad
\delta \theta = - \varepsilon.
\ea
It is easy to verify that the Hamiltonian, identified as being the new presymplectic potential ${\tilde U}^{(0)}$, is invariant under these infinitesimal gauge transformation above, namely, $\delta {\tilde U}^{(0)} = 0$.

We would like also to demonstrate that the anomaly was canceled. 
It will be carried out from Dirac's point of view. From the Lagrangian in Eq. (\ref{00390}) the sets of primary constraints are computed as, $\varphi_1 = \pi_0$, and $\chi_1 \,=\, - \pi_\theta  + \Psi$.
The primary Hamiltonian is ${\tilde U}^{(0)}_{primary} = {\tilde U}^{(0)} + \lambda_1\varphi_1 + \gamma_1\chi_1$.
Since the constraint $\varphi_1$ has no time evolution, the following secondary constraint is calculated as $\varphi_2 =\Omega_1 - q^2(a-1)\theta$, and no more constraints arise from the temporal stability condition. In this way, the total Hamiltonian is
\be
\label{00407a3}
{\tilde U}^{(0)}_{total} = {\tilde U}^{(0)} + \lambda_1\varphi_1 + \lambda_2\varphi_2 + \gamma_1\chi_1.
\ee

\noindent The time stability condition for the constraint $\chi_1$ allows us to determine the Lagrange multiplier $\lambda_3$. In this way,
the gauge invariant version of the model has three constraints, namely, $\varphi_1$, $\varphi_2$ and $\chi_1$. The corresponding Dirac
matrix, given by,
\be
\label{00407a4}
C(x - y) = \left( \begin{array}{ccc}
0 & -q^2(a-1) & 0 \\
q^2(a-1) & 0 & q^2(a-1) \\
0 & -q^2(a-1) & 0 \\
\end{array}\right) \delta(x-y),
\ee
is singular. As the Dirac matrix is singular, the model has both first and second-class constraints. Through a constraint
combination, we obtain a set of first-class constraints such as, $\tilde\chi_1 = - \pi_{\theta}  + \Psi - \pi_0$,
and a set of second-class constraints, given by $\tilde\varphi_1 = \varphi_1$ and $\tilde\varphi_2=\Omega_1 - q^2(a-1)\theta$.
It is again easy to verify that $\tilde\chi_1$ is a first-class constraint, identified as the Gauss law, while the others are second-class constraints. Note that the anomaly was removed. Hence, the Gauss law is also recognized as being the generator of the gauge transformation given in Eq. (\ref{00395}).

The model has one first-class and two second-class constraints and the phase space dimensions result in eight dependent fields, i.e., $(\phi,\pi_\phi, A_1,\pi_1,A_0,\pi_0,\theta,\pi_\theta)$. The first-class constraint eliminates two fields, while the second-class constraints eliminate two fields.  Hence, the model has four independent fields, i.e., there are two independent degrees of freedom.

In order to obtain the Dirac brackets, the set of second-class constraints, $\tilde\varphi_1$ and $\tilde\varphi_2$ , will be assumed equal to zero in a strong way. The Dirac brackets among the phase space fields are obtained as
\ba
\label{00408a1}
\lbrace\phi(x),\pi_\phi(y)\rbrace^* &=& \delta(x - y),\nonumber\\
\lbrace\phi(x),A_0(y)\rbrace^* &=& - \frac{1}{q(a-1)}\delta(x - y),\nonumber\\
\lbrace\pi_\phi(x),A_0(y)\rbrace^* &=& \frac{1}{q(a-1)}\partial_y\delta(x - y),\nonumber\\
\lbrace A_1(x),A_0(y)\rbrace^* &=& -\frac{1}{q^2(a-1)}\partial_y\delta(x - y),\nonumber\\
\lbrace A_1(x),\pi_{1}(y)\rbrace^* &=& \delta(x - y),\nonumber\\
\lbrace\pi_1(x),A_0(y)\rbrace^* &=&  \frac{1}{(a-1)}\delta(x - y),\nonumber\\
\lbrace\theta(x),\pi_\theta(y)\rbrace^* &=& \delta(x - y), \nonumber\\
\lbrace\pi_\theta(x),A_0(y)\rbrace^*  &=& \delta(x - y),\nonumber\\
\ea

\noindent the others are zero.  Note that the Dirac brackets among the original phase space fields were obtained before.
After this process, the model now have only one first-class constraint,  
$\chi=\tilde\chi_1|_{\tilde\varphi_1 = \tilde\varphi_2 = 0} = - \pi_\theta  + \Psi$ 
identified as the Gauss law, that satisfies the Poisson algebra, $\lbrace\chi(x),\chi(y)\rbrace^* = 0$.
In this way, the anomaly was eliminated, the symmetry is preserved, and the fundamental brackets among the original phase space fields were
reobtained. Note that the Gauss law is the generator of the gauge symmetry given in (\ref{00395}).

Once more, the number of the independent degrees of freedom matches with the result obtained in the second-class case. The invariant
model has a phase space $(\phi,\pi_\phi,A_1,\pi_1,\theta,\pi_\theta)$, with six dependent fields, and has a first-class constraint
which eliminates two fields.  Consequently, the model has two independent degrees of freedom.

The remaining symmetry will be eliminated with the introduction of the unitary gauge-fixing term, given by $\theta = 0$.
Hence, both the noninvariant Hamiltonian and the corresponding Dirac brackets computed in the beginning of this section are reobtained, then recovering the anomaly. In this way, we conclude that the new symplectic gauge-invariant formalism does not change the physics contents inside the model.

\section{The non-Abelian extension of the Proca model}
\renewcommand{\theequation}{6.\arabic{equation}}
\setcounter{equation}{0}

The non-Abelian extension of the Proca model has its dynamics governed by the following Lagrangian density\footnote{The BFFT embedding for the non-Abelian Proca theory was discussed in \cite{bb100}},
\be
\label{N0000}
{\cal L} = -\frac 14 F_{\mu\nu}^a\,F_a^{\mu\nu} + \frac 12 m^2\,A_\mu^a A_a^\mu,
\ee
with $F_{\mu\nu}^a = \partial_\mu A_\nu^a - \partial_\nu A_\mu^a + g C^a_{bc}A_\mu^bA_\nu^c$.
The antisymmetric tensor $C^a_{bc} \: (C^a_{bc} = - C^a_{cb})$, are in fact a set of real constants, known as the structure constants of the gauge group, and satisfy the following property, 
\be
\label{N0020}
C^a_{bc}C^d_{ae} + C^a_{eb}C^d_{ac} + C^a_{ce}C^d_{ab} = 0.
\ee

The first-order form follows,
\ba
\label{N0030}
{\cal L}^{(0)} &=& \pi_a^i {\dot A}^a_i - \frac 12 (\pi_a^i)^2 + A_0^a\Omega_a - \frac 12 m^2 A_i^aA^i_a \nonumber \\
&-& \frac 12 m^2 A_0^aA^0_a - \frac 14 F_{kj}^aF_{kj}^a,
\ea
where $\Omega_a = \partial_i\pi^i_a - gC^b_{ca} \pi_b^iA_i^c + m^2A_a^0$.  The presymplectic variables are given by
$\xi_\alpha^a = (A_i^a,\pi_i^a,A_0^a)$ and the presymplectic matrix is
\ba
f^{(0)} &=&
\begin{pmatrix}
0 & - \delta_{ji}\delta^{ba} & 0
\cr \delta_{ij}\delta^{ab} & 0  & 0
\cr 0 & 0 & 0
\end{pmatrix}
\delta^{(3)}(\vec x - \vec y).
\ea

\noindent Since this matrix is singular, it has a zero-mode that generates the constraint $\Omega_a$. The first-order Lagrangian through a Lagrange multiplier, is,
\ba
\label{N0060}
{\cal L}^{(1)} &=& \pi_a^i {\dot A}^a_i + \Omega_a\dot\eta^a - \frac 12 (\pi_a^i)^2 + A_0^a\Omega_a - \frac 12 m^2 A_i^aA^i_a  \nonumber \\
&-&
\frac 12 m^2 A_0^aA^0_a - \frac 14 F_{kj}^aF_{kj}^a.
\ea
The new group of presymplectic variables is $\xi_\alpha^a = (A_i^a,\pi_i^a,A_0^a,\eta^a)$, and the new presymplectic matrix is
\ba
\label{N0070}
f^{(1)} &=&
\begin{pmatrix}
0 & - \delta_{ji}\delta^{ba} & 0 & - g C^{ab}_d\pi^d_i(y)\cr \delta_{ij}\delta^{ab} & 0  & 0 & D^{ba}(y)
\cr 0 & 0 & 0 & m^2 \delta^{ab} \cr g C^{ba}_d\pi^d_j(x) & - D^{ba}(x) &
- m^2\delta^{ba} & 0
\end{pmatrix} \nonumber \\
&\cdot& \delta^{(3)}(\vec x - \vec y)\,\,,
\ea
where $D^{ab}(w)=\delta^{ab}\partial_i^w - g C^{ab}_d A^d_i(w)$.  This matrix is nonsingular. The model now can be reformulated as a gauge invariant field theory. 


The first-order Lagrangian (\ref{N0030}) can be rewritten as
\ba
\label{N0090}
{\tilde {\cal L}}^{(0)} &=& \pi_a^i {\dot A}^a_i + \Psi_a\dot\theta^a - \frac 12 (\pi_a^i)^2 + A_0^a\Omega_a   \\
&-& \frac 12 m^2 A_i^aA^i_a - \frac 12 m^2 A_0^aA^0_a - \frac 14 F_{kj}^aF_{kj}^a - G, \nonumber 
\ea

\noindent where the arbitrary functions are
\ba
\label{N0100}
\Psi_a &\equiv& \Psi_a(A_i^a,\pi_i^a,A_0^a,\theta^a), \\
G &\equiv& G (A_i^a,\pi_i^a,A_0^a,\theta^a)=\sum_{n=0}^\infty{\cal G}^{n}(A_i^a,\pi_i^a,A_0^a,\theta^a)\,\,. \nonumber 
\mbox{}
\ea

\noindent The $G$ function obeys a boundary condition given by,
\ba
\label{N0110}
G \equiv (A_i^a,\pi_i^a,A_0^a,\theta^a=0)={\cal G}^{0}(A_i^a,\pi_i^a,A_0^a,\theta^a=0)=0. \nonumber \\
\mbox{}
\ea
In this context, the corresponding presymplectic matrix is
\be
\label{N0130}
f^{(0)} =
\begin{pmatrix}
0 & - \delta_{ji}\delta^{ba} & 0 & \frac{\partial\Psi_b(y)}{\partial A^a_i(x)}
\cr \delta_{ij}\delta^{ab} & 0  & 0 & \frac{\Psi_b(y)}{\partial \pi^a_i(x)}
\cr 0 & 0 & 0 & \frac{\partial\Psi_b(y)}{\partial A^a_0(x)}
\cr - \frac{\partial\Psi_a(x)}{\partial A^b_j(y)} & - \frac{\partial\Psi_a(x)}{\partial \pi^b_j(y)} & - \frac{\partial\Psi_a(x)}{\partial A^b_0(y)} & 0
\end{pmatrix}
\delta^{(3)}(\vec x - \vec y).
\ee

In order to determine the functions $\Psi_a$, we analyze the symmetry related to the following zero-mode,
$\bar\nu^{(0)} =
\begin{pmatrix}
\partial^x_i & 0 & 0 & 1
\end{pmatrix}$
which produces a set of differential equations which allows us to compute the $\Psi_a$ function as $\Psi_a =  - \partial_i \pi^i_a(x)$.
Consequently, the first-order Lagrangian can be rewritten as
\be
\label{N0160}
{\tilde {\cal L}}^{(0)} = \pi_a^i {\dot A}^a_i - (\partial_i \pi^i_a)\dot\theta^a - {\tilde V}^{(0)},
\ee
where the presymplectic potential is
\ba
\label{N170}
{\tilde V}^{(0)} &=& \frac 12 (\pi_a^i)^2 - A_0^a\Omega_a + \frac 12 m^2 A_i^aA^i_a + \frac 12 m^2 A_0^aA^0_a \nonumber \\
&+& \frac 14 F_{kj}^aF_{kj}^a + G.
\ea

For the hidden symmetry inside the model, we have,
\be
\label{N0180}
\int_x \bar\nu^{(0)}_\alpha(w)\frac{\partial{\tilde V}(x)}{\partial \xi^a_\alpha(w)} = 0.
\ee

\noindent The linear correction term in $\theta$ is given by, 
\be
\label{N0190}
\int_x \left\{\partial^w_i\frac{\partial V(x)^{(0)}}{\partial A_i^f(w)} + \frac{{\cal G}^{(1)}(x)}{\partial \theta^f(w)}\right\} = 0.
\ee
After an integration we have that
\ba
\label{N0200}
&&{\cal G}^{(1)}(x) \nonumber \\
&=&-\, g C_{fa}^b \partial^x_i(A_0^a(x)\pi_b^i(x))\theta^f(x) - m^2 (\partial^x_iA^i_f)\theta^f(x)\nonumber\\
 &-& \frac 12 \int_y \partial^y_i \left(F^a_{kj}(x)\frac{\partial F_a^{kj}(x)}{\partial A_i^f(y)}\right)\theta^f(y).
\ea
Now, we will compute the quadratic term, namely,
\be
\label{N0210}
\int_x \left\{\partial^w_i\frac{\partial {\cal G}^{(1)}(x)}{\partial A_i^f(w)} + \frac{{\cal G}^{(2)}(x)}{\partial \theta^f(w)}\right\} = 0.
\ee
Integrating this relation in $\theta^f(w)$, the quadratic correction term is obtained as
\ba
\label{N0220}
{\cal G}^{(2)}(x) &=&  \frac 12 m^2 (\partial_x^i \theta^f(x))^2 \nonumber \\
&+&
\frac 12\int_{\theta^f(x)}  \int_w \partial^w_i \int_y \left[(\partial^y_l{\cal A}^{il}_{fb})\theta^b(y)\right], \nonumber \\
\mbox{}
\ea
where
\ba
\label{N0230}
{\cal A}^{il}_{fb} = \frac{\partial F^a_{kj}(x)}{\partial A_i^f(w)} \frac{\partial F_a^{kj}(x)}{\partial A_l^b(y)} +
F^a_{kj}(x)\frac{\partial^2 F_a^{kj}(x)}{\partial A_i^f(w)\partial A_l^b(y)}. \nonumber \\
\mbox{}
\ea

In this way, two correction terms as functions of $\theta_a$ $({\cal G}^{(3)}(x)$ and ${\cal G}^{(4)}(x))$ were computed yet.  Let us compute the first one. It can be carried out from the following relation,
\ba
& &\int_z\left\{\partial_n^z \left[\frac 12 \int_{\theta^f(x)}\int_w \partial_k^w \int_y \partial^y_l\frac{\partial{\cal A}^{kl}_{fb}}{\partial A_n^g(z)} \theta^b(y)\right] \right. \nonumber \\
&+& \left. \frac{{\cal G}^{(3)}(x)}{\partial \theta_g(z)} \right\} = 0 \nonumber
\ea
and we can write that,
\ba
\label{N0240}
& &{\cal G}^{(3)}(x) \\
&=& - \frac 12 \int_{\theta^g(z)} \int_z \partial_n^z \int_{\theta_f(x)} \int_w \partial_k^w \int_y \partial^y_l\frac{\partial{\cal A}_{fb}^{kl}}{\partial A^n_g(z)} \theta^b(y). \nonumber 
\mbox{}
\ea
\noindent Finally, the last correction term is,
\ba
\label{N0250}
&&{\cal G}^{(4)}(x) = \frac 12 \int_{\theta^h(v)} \int_v \partial_i^v \int_{\theta_g(z)} \int_z \partial_n^z \int_{\theta_f(x)} \cdot \nonumber \\
&& \cdot \int_w
\partial_k^w \int_y \partial^y_l\frac{\partial^2{\cal A}_{fb}^{kl}}{\partial A^i_h(v)\partial A^n_g(z)} \theta^b(y).
\ea
Therefore, the gauge invariant first-order Lagrangian is
\be
\label{N0260}
{\tilde {\cal L}}^{(0)} = \pi_a^i {\dot A}^a_i - (\partial_i \pi^i_a)\dot\theta^a - {\tilde V}^{(0)},
\ee
where the presymplectic potential, identified as being the gauge invariant Hamiltonian, is given by
\ba
\label{N270}
{\tilde V}^{(0)} &=& \frac 12 (\pi_a^i)^2 - A_0^a\Omega_a + \frac 12 m^2 (A_i^a)^2 \nonumber \\
&+& \frac 12 m^2 (A_0^a)^2 + \frac 14 (F_{kj}^a)^2 \\
&-& g C_{fa}^b \partial^x_i(A_0^a(x)\pi_b^i(x))\theta^f(x)\nonumber\\ 
&-& m^2 (\partial^x_iA^i_f)\theta^f(x) + \frac 12 m^2 (\partial^i \theta^f(x))^2 \nonumber \\
&+& I_1 + I_2 + I_3 + I_4 \nonumber 
\ea
where
\be
I_1 = - \frac 12 \int_y \partial^y_i \left(F^a_{kj}(x)\frac{\partial F_a^{kj}(x)}{\partial A_i^f(y)}\right)\theta^f(y)\nonumber 
\ee
\be
I_2 = + \frac 12 \theta_f(w) \int_w \partial^w_i \int_y \left[(\partial^y_l{\cal A}_{fb}^{il})\theta^b(y)\right] \nonumber
\ee
\be
I_3 = - \frac 12 \int_{\theta^g(z)} \int_z \partial_n^z \int_{\theta_f(x)} \int_w \partial_k^w \int_y \partial^y_l\frac{\partial{\cal A}_{fb}^{kl}}{\partial A^n_g(z)} \theta^b(y) \nonumber 
\ee
\be
I_4 = - \frac 12 \int_{\theta^h(v)} \int_v \partial_i^v \int_{\theta_g(z)} \int_z \partial_n^z \int_{\theta_f(x)}  \nonumber 
\ee
\be
\cdot \int_w \partial_k^w \int_y \partial^y_l\frac{\partial^2{\cal A}_{fb}^{kl}}{\partial A^i_h(v)\partial A^n_g(z)} \theta^b(y)\,\,. \nonumber
\ee

The second-order Lagrangian is
\ba
\label{secondVI}
&&{\tilde {\cal L}}^{(0)} = \frac 12 \(\dot{A}_i^a\)^2 + 2 \dot{A}_i^a \p^i \dot{\theta}^a + \dot{A}_i^a \p^i A_{0a} + \(\p_i \dot{\theta}^a \)^2 \nonumber \\
&-& \frac 32 \(\p_i A^a_0 \)^2 - \dot{A}_i^a \p^i \theta_a - \p_i A^a_0 \p^i \dot{\theta}_a \nonumber \\
&+& g C^a_{bc} \p^i A_{0a} A^b_i A^c_0 - 3 g C^a_{bc} \p^i A_{0a} \p_i \theta^b A^c_0 - g C^a_{bc} \dot{A}_a^i A^b_i A^c_0 \nonumber \\
&-& g C^a_{bc} \p^i \dot{\theta}_a A^b_i A^c_0 + \frac 12 g^2 \( C^a_{bc} A^b_i A^c_0 \)^2 
- \frac 32 g^2 \( C^a_{bc} \p_i \theta^b A^c_0 \)^2 \nonumber \\
&-& g C^a_{bc} A^c_0 \p^i \theta^b \( \dot{A}_i^a + \p_i \dot{\theta}_a \) 
+ g^2 C^a_{bc} C_{amn} \p_i \theta^b A^c_0 A^n_0 A^{im} \nonumber \\
&+& \frac 12 m^2 (A^a_0)^2 - \frac 12 m^2 (A^a_i)^2 - \frac 14 \( F^a_{ij} \)^2 \\
&+& m^2 \( \p_i A^i_a \) \theta^a - \frac 12 m^2 \( \p^i \theta^a \)^2 - I_1 -I_2 - I_3 - I_4 \nonumber
\ea

Let us make an analysis using the Dirac point of view.  We start with the set of primary constraints, $\Omega_1^a = \partial^i\pi_i^a + \pi_\theta^a$ and
$\chi_1^a= \pi_0^a$.
For the first set of constraints, the time stability condition is satisfied $(\dot\Omega_1^a =0)$.  For the second one, the following secondary constraints are required, $\chi_2^a = \Omega^a - g C_f^{ba}\pi_b^i\partial_i\theta^f$.
Hence, the total Hamiltonian is
${\cal H} = {\tilde{\cal H}} + \lambda_a^1\Omega^a_1 + \zeta_a^1\chi_1^a + \zeta_a^2\chi_2^a$,
where $\lambda_a^1$, $\zeta_a^1$ and $\zeta_a^2$ are Lagrange multipliers. Since the Poisson brackets among those constraints are
\ba
\label{N0310}
&&\lbrace\Omega^a_1(x),\Omega^b_1(y)\rbrace = \:\:
\lbrace\Omega^a_1(x),\chi^b_1(y)\rbrace = 0,\nonumber \\
&&\lbrace\Omega^a_1(x),\chi^b_2(y)\rbrace = 0,\:\;
\lbrace\chi^a_1(x),\chi^b_2(y)\rbrace = -m^2\delta^{ab}\delta^{(3)}(\vec x - \vec y),\nonumber\\
&&\lbrace\chi^a_2(x),\chi^b_2(y)\rbrace = 2gC_{d}^{ab} \chi_2^d(x)\delta^{(3)}(\vec x - \vec y) \nonumber \\
&& \qquad \qquad \qquad \qquad - 2 gm^2C_{d}^{ab}A_0^d(x)\delta^{(3)}(\vec x - \vec y),\nonumber
\ea
no more constraints arise. Notice that some brackets above are zero, indicating that there are both first and second-class constraints.
Indeed, the first-class constraint is $\Omega^a_1$ and the second-class are $\chi^a_1$ and $\chi^a_2$. In agreement with Dirac's
procedure, the second-class constraints can be taken equal to zero in a strong way.  This allows us to compute the primary Dirac brackets.
Due to the Maskawa-Nakajima theorem \cite{NM}, the primary Dirac brackets among the phase space fields are canonical. To demonstrate this, the
brackets are computed explicitly. The Dirac matrix is
\be
\label{N0320}
C =
\begin{pmatrix}
0 & -m^2\delta^{cd} \cr m^2\delta^{dc} & B^{cd}
\end{pmatrix}
\delta^{(3)}(\vec x - \vec y),
\ee
with $B^{cd} = 2gC_{b}^{cd} \chi_2^b(x)- 2 gm^2C_{b}^{cd}A_0^b(x)$.
The inverse of the Dirac matrix is
\be
\label{N0335}
C^{(-1)} = \frac {1}{m^2}
\begin{pmatrix}
\frac {B^{cd}}{m^2} & \delta^{cd} \cr - \delta^{dc} & 0
\end{pmatrix}
\delta^{(3)}(\vec x - \vec y).
\ee
In accordance with Dirac's process, the Dirac brackets among the phase space fields are obtained as
\ba
\label{N0340}
\lbrace A^a_i(x), \pi^b_j(y)\rbrace^* &=& \delta^{ab}\delta^{(3)}(\vec x - \vec y), \nonumber\\
\lbrace A^a_i(x), A^b_0(y)\rbrace^* &=& -\frac {1}{m^2}\partial^x_i\delta^{(3)}(\vec x - \vec y) \nonumber \\
&+& \frac {g}{m^2} C^{ab}_f A_i^f(x)\delta^{(3)}(\vec x - \vec y), \nonumber\\
\lbrace \pi^a_i(x), A^b_0(y)\rbrace^* &=& -\frac{1}{m^2} g C^{ab}_e\pi^e_i\delta^{(3)}(\vec x - \vec y),\nonumber\\
\lbrace A^a_0(x), A^b_0(y)\rbrace^* &=& -\frac{g}{m^2} C_e^{ab}A^e_0(x)\delta^{(3)}(\vec x - \vec y), \nonumber \\
\lbrace A^a_0(x), \pi_\theta^b(y)\rbrace^* &=& \frac {g}{m^2} C_e^{ab}\partial^x_i\pi^e_(x)\delta^{(3)}(\vec x - \vec y),\\
\lbrace \theta^a(x), \pi_\theta^b(y)\rbrace^* &=& \delta^{ab}\delta^{(3)}(\vec x - \vec y),\nonumber
\ea

\noindent and the others are zero.  Finally, the infinitesimal gauge transformations are obtained, namely,

\ba
\label{N0350}
\delta A_i^a &=& - \partial_i^x\varepsilon^a,\nonumber\\
\delta \pi_i^a &=& 0,\nonumber\\
\delta A_0^a &=& 0,\\
\delta \theta^a &=& \varepsilon^a,\nonumber\\
\delta \pi_\theta^a &=& 0,\nonumber
\ea
which lead us to the invariant Hamiltonian.

To demonstrate that the gauge invariant formulation of the non-Abelian Proca model is dynamically equivalent to the original noninvariant
model, the symmetry is fixed by using the unitary gauge fixing procedure, $\varphi^a = \theta^a\approx 0$,
which leads to the bracket below, $\lbrace \Omega^a_1(x),\varphi^b(y)\rbrace = - \delta^{ab}\delta^{(3)}(\vec x - \vec y)$.
So, a new Dirac brackets must be computed. The corresponding Dirac matrix for this set of constraints is
\be
\label{N0380}
C =
\begin{pmatrix}
0 & -1 \cr 1 & 0
\end{pmatrix}
\delta^{(3)}(\vec x - \vec y).
\ee
Using the inverse of this matrix, the Dirac brackets among the physical phase space fields can be computed, which is equal to the one calculated from the original description.
This result demonstrate again that the symplectic formalism can be seen in fact as a mapping between the original theory and the final one.


\section{Hidden symmetries of a fluid dynamical model}
\renewcommand{\theequation}{7.\arabic{equation}}
\setcounter{equation}{0}

It was demonstrated in \cite{bh} that the relativistic theories of membranes are integrable systems through the transformation of the problem into a two-dimensional fluid dynamics one.  In this case, the potential term is proportional to the inverse of mass density, i.e., $V\propto \rho^{-1}$.
This subject is connected in some way to other fields like the parton model \cite{jevicki}, hydrodynamical description of quantum mechanics \cite{mm}, black hole cosmology \cite{kmp} and hydrodynamics of superfluid systems \cite{schakel}.

The majority of these cases aimed to find the solutions of the Galileo invariant system in $d$-dimensions in connection with the solutions of the relativistic $d$-brane system in $(d+1)$-dimension \cite{jackiw}.

Let us start with the linear (or nonlinear), Schr\"odinger Lagrangian theory defined in a 
$d$-dimensional ($\bf r)$ space evolving in time $(t)$,
\be
\label{h1}
L_S\,=\,\int d^d {\bf r} \big[ i \Psi^*\,\dot{\Psi}\,-\,\frac 12\,(\nabla\Psi^*)\cdot(\nabla\Psi)\,-\,\nabla(\Psi^*\,\Psi) \big]
\ee
where $\bar{V}$ represents any nonlinear interaction and 
\be
\Psi\,=\,\rho^{1/2}\,e^{i\theta}\,\,,
\ee
where $\rho\equiv\rho(t,{\bf r})$ and $\theta\equiv\theta(t,{\bf r})$ is the velocity potential \cite{schakel} into the Schr\"odinger Lagrangian.  Now we can write the fluid dynamical model
\be
\label{h2}
L\,=\,\int d^d {\bf r} \big[ \theta\,\dot{\rho}\,-\,\frac 12\,\rho\,\nabla\theta \cdot \nabla \theta\,-\,V(\rho) \big]
\ee
where $V(\rho) = \bar{V}(\rho)\,+\, \frac{1}{8} \frac{(\nabla \rho)^2}{\rho}$, which is the hydrodynamical form of the Schr\"odinger theory (details can be found in \cite{01}).  We can say here that the connection between the fluid model, the membrane and its generalization to $d$-brane systems only appears under the very specific density-dependent interaction potential $V\,=\,g/\rho$ \cite{jackiw}.  The symmetries and the corresponding generators for the fluid dynamics model can be found in \cite{01}.

The Lagrangian in (\ref{h2}) is already written in first-order form, so we can write ${\cal L}^{(0)}\,=\,\theta\,\dot{\rho}\,-\,V^{(0)}$ where $V^{(0)}\,=\,\frac 12\, \rho\,\partial_i \theta\,\partial^i\,\theta\,+\,V(\rho)$.  
The symplectic coordinates are $\xi^{(0)\beta}\,=\,(\rho,\theta)$, $A^{(0)}_{\rho}\,=\,\theta$ and $A_\theta^{(0)}\,=\,0$.

The zeroth-iterative symplectic matrix
\be
f^{(0)}\,=\,\left(
\begin{array}{cc}
0 & -\delta^{(d)}({\bf r}-{\bf r}')  \\
\delta^{(d)}({\bf r}-{\bf r}') & 0  
\end{array}
\right) 
\ee
is non-singular and the model is not gauge invariant.

We have that,
\be
\tilde{\cal L}^{(0)}\,=\,\theta\,\dot{\rho}\,+\,\Psi\,\dot{\eta}\,-\,\tilde{V}^{(0)}
\ee
where $\tilde{V}^{(0)}\,=\,\frac 12\,\rho\,\partial_i\,\theta\,\partial^i\,\theta\,+\,V(\rho)$,
and $\Psi\equiv\Psi(\rho,\theta)$ and $G\equiv\,G(\rho,\theta,\eta)$ and the symplectic matrix is 
\be
\label{h3}
f^{(0)}\,=\,\left(
\begin{array}{ccc}
0 & -\delta^{(d)}({\bf r}-{\bf r}') & \frac{\delta\,\Psi_{{\bf r}'}}{\delta\,\rho({\bf r})}   \\
\delta^{(d)}({\bf r}-{\bf r}') & 0 & \frac{\delta\,\Psi_{{\bf r}'}}{\delta\,\theta({\bf r})}  \\
-\frac{\delta\,\Psi_{{\bf r}}}{\delta\,\rho({\bf r}')} & -\frac{\delta\,\Psi_{{\bf r}}}{\delta\,\theta({\bf r}')} & 0
\end{array}
\right) 
\ee
where $\Psi_{\bf r}\equiv\Psi(\rho({\bf r})),\theta({\bf r}))$, 
$\Psi_{{\bf r}'}\equiv\Psi(\rho({\bf r}'),\theta({\bf r}'))$ and the zero-mode 
${\tilde{\nu}}^{(0)}({\bf r})$ satisfies the relation
\be
\label{h4}
\int d^d {\bf r}\, \tilde{\nu}^{(0)\tilde{\theta}}\,({\bf r})\,\tilde{f}_{\tilde{\theta}\tilde{\beta}}({\bf r},{\bf r}')
\ee
which provides us with the set of equations that allows the determination of $\Psi$ explicitly.  

The WZ gauge symmetry is related to the following zero-mode $\tilde{\nu}^{(0)}\,=\,(1\:\:1\:\:-1)$
since this zero-mode and the symplectic matrix (\ref{h3}) must satisfy the gauge symmetry condition in (\ref{h4}).  We can obtain the set of equations
\ba
\int d^d {\bf r} \Big( \delta^{(d)}\,({\bf r}-{\bf r}')\,+\,\frac{\delta\,\Psi_{{\bf r}}}{\delta\,\rho({\bf r}')} \Big) &=&0 \nonumber \\
\int d^d {\bf r} \Big( -\delta^{(d)}\,({\bf r}-{\bf r}')\,+\,\frac{\delta\,\Psi_{{\bf r}}}{\delta\,\theta({\bf r}')} \Big) &=&0 \\
\int d^d {\bf r} \Big( \frac{\delta\,\Psi_{{\bf r}}}{\delta\,\rho({\bf r}')}\,+\,\frac{\delta\,\Psi_{{\bf r}}}{\delta\,\rho({\bf r}')} \Big) &=&0\,\,. \nonumber 
\ea
After an integration, $\Psi$ can be written as $\Psi({\bf r})\,=\,\theta({\bf r})\,-\,\rho({\bf r})$
and the symplectic matrix,
\be
\tilde{f}^{(0)}\,=\,\left(
\begin{array}{ccc}
0 & -1 & -1 \\
1 & 0 & 1 \\
1 & -1 & 0 
\end{array}
\right) \delta^{(d)}\,({\bf r}-{\bf r}')
\ee
which is singular.  Hence, the first-order Lagrangian is 
\be
\label{XXXXX}
\tilde{\cal L}^{(0)}\,=\,\theta\,\dot{\rho}\,+\,(\theta\,-\,\rho)\,\dot{\eta}\,-\,\tilde{V}^{(0)}\,\,,
\ee
where 
$\tilde{V}^{(0)}\,=\,\frac 12\,(\rho\,+\,\eta)\,(\partial_i\,\theta)^2\,+\,V(\rho-\eta)$.
The zero-mode $\tilde{\nu}^{(0)}$ is the generator of infinitesimal gauge transformation, then 
\ba
\delta\,\rho({\bf r},t) &=&\varepsilon({\bf r}',t)\,\delta^{(d)}({\bf r}-{\bf r}') \nonumber \\
\delta\,\theta({\bf r},t) &=&\varepsilon({\bf r}',t)\,\delta^{(d)}({\bf r}-{\bf r}') \\
\delta\,\eta({\bf r},t) &=&-\varepsilon({\bf r}',t)\,\delta^{(d)}({\bf r}-{\bf r}') \nonumber 
\ea
where $\varepsilon({\bf r},t)$ is an infinitesimal time-dependent parameter. It was shown in \cite{01} that
the Lagrangian density in (\ref{XXXXX}) becomes,
\be
\tilde{\cal L}\,=\,-(\rho\,-\,\eta)\,\dot{\theta}\,-\,\frac 12\,(\rho\,-\,\eta)\,(\partial_i\,\theta)^2\,+\,V(\rho-\eta)\,\,,
\ee
which is the same result obtained in \cite{nb}.  This Lagrangian can also be written as
\ba
\tilde{\cal L}&=&(\rho - \eta )\,\frac{\p V(\rho-\eta)}{\p \eta} + V(\rho-\eta) \qquad \mbox{or} \nonumber \\
\tilde{\cal L}&=&-\,(\rho - \eta )\,\frac{\p V(\rho-\eta)}{\p \rho} + V(\rho-\eta)\,\,,
\ea
which depends on both equations of motion for $\eta$ and for $\rho$ respectively.

As a second example \cite{01} of zero-mode for this theory we have that the zero-mode $\tilde{\nu}^{(0)}\,=\,(1\:\:0\:\:-1)$ will generate the gauge-invariant first-order Lagrangian \cite{01},
\be
\tilde{\cal L}\,=\,-(\rho\,+\,\eta)\,\dot{\theta}\,-\,\frac 12\,(\rho\,+\,\eta)\,(\partial_i\,\theta)^2
\,+\,V(\rho\,+\,\eta)\,\,,
\ee
and the infinitesimal gauge transformations will be obtained as
\ba
\delta\,\rho({\bf r},t) &=&\varepsilon({\bf r}',t)\,\delta^{(d)}({\bf r}-{\bf r}')\,\,, \qquad \delta\,\theta({\bf r},t) = 0 
\nonumber \\
\delta\,\eta({\bf r},t) &=&-\varepsilon({\bf r}',t)\,\delta^{(d)}({\bf r}-{\bf r}')  
\ea

In \cite{01}, the interested reader can obtain other six symmetries within the fluid model described 
in (\ref{h1}).


\section{Hidden symmetry in the rotational fluid model}
\renewcommand{\theequation}{8.\arabic{equation}}
\setcounter{equation}{0}

In \cite{01}, some of us have demonstrated that the irrotational fluid model has a set of dynamically equivalent WZ gauge invariant versions. Also, the extra global symmetries, namely, Galileo antiboost and time rescaling, first obtained in \cite{02}, were promoted to local symmetries.

In this section, we will investigate the symmetries of the rotational fluid model, but now they have an extra term, like $k\rho\(\partial_i\theta+\alpha\partial_i\beta \)^2$, where $k$ is a constant, $\rho\equiv \rho(t,\vec r)$ is the mass density and $\theta\equiv \theta(t,\vec r)$ is the velocity potential. The terms introduce a dissipative force into the model. 

To establish our ideas, we will investigate how the inclusion of a dissipative term affects the dynamics of the fluid model considering only one of these zero modes introduced in \cite{01}. 

It is well known that systems that have vorticity and/or viscosity have Casimir invariants which obstruct the construction of a canonical formalism for a fluid, as demonstrated in \cite{RJ}. However, this obstruction can be eliminated using the Clebsch parameters, as it was shown by Lin \cite{Lin} and by two of us in \cite{NW2}. In fact, with the introduction of Clebsch parameters, it is possible to obtain a Lagrangian density for the rotational fluid with dissipation, 3-dimensional, as being
\be
\label{002}
{\cal L} = -\rho(\dot \theta + \alpha\dot\beta) - V,
\ee
where the presymplectic potential is
\be
\label{03}
V =\frac {1}{2}(1-k)\rho (\partial_i \theta+ \alpha\partial_i \beta) (\partial^i\theta+ \alpha\partial^i\beta) + V(\rho).
\ee
The presymplectic coordinates are $\xi^{(0)}=(\rho, \theta, \alpha, \beta)$ and the corresponding zeroth-iterative one-form canonical momenta is given by $A_{\rho}^{(0)} = 0, \:
A_{\theta}^{(0)} = -\rho, \:A_{\alpha}^{(0)} = 0, \: A_{\beta}^{(0)} = - \alpha\rho$. The zeroth-iteration presymplectic matrix, given by
\begin{equation}
f^{(0)} = \left(
\begin{array}{cccc}
0           & -\delta(\vec r -\vec r^{\prime}) & 0 & -\alpha\\
\delta(\vec r -\vec r^{\prime})&         0     & 0 & 0\\
0 &         0      &0 & -\rho\\
\alpha & 0 & \rho & 0 
\end{array}
\right),
\end{equation}
is a nonsingular matrix.  The  model is not gauge-invariant. 

Although the symplectic formalism does not restrain the dimension of the model, we choose a 3-dimensional description for the rotational fluid to place our work in a correct perspective in comparison with others. 

The first-order Lagrangian ${\cal L}^{(0)}$, equation (\ref{002}), with additional arbitrary terms $(\Psi, G)$ is given by
\begin{equation}
\label{06}
{\tilde{\cal L}}^{(0)} = -\rho(\dot \theta + \alpha\dot\beta) + \Psi\dot\eta - {\tilde V}^{(0)},
\end{equation}
where
\be
\label{07}
{\tilde V}^{(0)} =\frac{1}{2}(1-k)\rho (\partial_i \theta + \alpha\partial_i \beta) (\partial^i\theta + \alpha\partial^i\beta) + V(\rho) + G,
\ee
and $\Psi\equiv\Psi(\rho,\theta)$ and $G\equiv G(\rho,\theta,\eta)$ are arbitrary functions that must be calculated. The presymplectic coordinates are ${\tilde\xi}^{(0)}=(\rho,\theta,\alpha,\beta,\eta)$ while the presymplectic matrix is
\be
\label{08}
{\tilde f}^{(0)} = \left(
\begin{array}{ccccc}
 0 & - \delta(\vec r - \vec r^{\prime}) &  0 & -\alpha & {\frac{\delta\Psi_{\vec r^{\prime}}}{\delta \rho(\vec r)}}\\ \nonumber
\delta(\vec r - \vec r^{\prime}) &  0 & 0& 0 & {\frac{\delta\Psi_{\vec r^{\prime}}}{\delta \theta(\vec r)}} \\ 
0& 0 & 0 & -\rho & {\frac{\delta\Psi_{\vec r^{\prime}}}{\delta \alpha(\vec r)}}\\

\alpha & 0 & \rho  & 0 & {\frac{\delta\Psi_{\vec r^{\prime}}}{\delta \beta(\vec r)}}\\

 - {\frac{\delta\Psi_{\vec r}}{\delta \rho(\vec r^{\prime})}} & - {\frac{\delta\Psi_{\vec r}}{\delta \theta(\vec r^{\prime})}} &  - {\frac{\delta\Psi_{\vec r}}{\delta \alpha(\vec r^{\prime})}}& - {\frac{\delta\Psi_{\vec r}}{\delta \beta(\vec r^{\prime})}}&0\nonumber
\end{array}\right),
\ee
where $\Psi_{\vec r} \equiv \Psi(\rho(\vec r),\theta(\vec r))$ and $\Psi_{\vec r^{\prime}} \equiv \Psi(\rho(\vec r^{\prime}),\theta(\vec r^{\prime}))$.

With a general zero-mode, ${\tilde\nu}^{(0)}=  \left(\begin{array}{ccccc}
a&b&c&d&-1
\end{array}\right)$,
we have the following set of differential equations
\ba
& &\int \,\, {\rm d}\vec r \,\,\left(b\delta(\vec r -\vec r^{\prime}) + \frac{\partial\psi}{\partial\rho} + d\alpha\delta(\vec r -\vec r^{\prime}) \right)
=0,\nonumber\\
& &\int \,\, {\rm d}\vec r \,\,\left(-a\delta(\vec r -\vec r^{\prime}) + \frac{\partial\psi}{\partial\theta} \right)=0,\\
& &\int \,\, {\rm d}\vec r \,\,\left(\frac{\partial\psi}{\partial\alpha} + d\rho\delta(\vec r -\vec r^{\prime})\right)=0,\nonumber\\
& &\int \,\, {\rm d}\vec r \,\,\left(\frac{\partial\psi}{\partial\beta} - a\alpha - c\rho\delta(\vec r -\vec r^{\prime})\right)=0.\nonumber
\ea
Hence we obtain that
\be
\Psi(\vec r) = -b\rho - \alpha d\rho +a\theta +\alpha a\beta +c\rho\beta.
\ee
In order to have a final solution, we consider that $a=c=0$, then
${\tilde\nu}^{(0)}=  \left(\begin{array}{ccccc}
0&1&0&1&-1
\end{array}\right)$, and $\Psi(\vec r) = -(1+\alpha)\rho$.

Then,
\ba
\label{015}
& &\int {\rm d} \vec r^{\prime} \;\;\tilde\nu^{(0)\tilde\beta}(\vec r)\;\frac{\delta {\tilde V}^{(0)}(\vec r^{\prime})}{\delta {\tilde\xi}^{\tilde\beta}(\vec r)} = 0\,\,,\nonumber\\
& &\int {\rm d} \vec r^{\prime} \;\;\Big\{(1+\alpha) \rho (\partial_i^{\prime}\theta+\alpha\partial_i^{\prime}\beta) \partial_i^{\prime}\delta(\vec r^{\prime} - \vec r)  \nonumber\\
&+& \sum_{n=1}\left(\frac{\partial {\cal G}^{(n)}}{\partial\theta}+\frac{\partial {\cal G}^{(n)}}{\partial\beta}-\frac{\partial {\cal G}^{(n)}}{\partial\eta}\right)\Big\}=0\,\,. 
\ea
To compute the first correction term as function of $\eta$, ${\cal G}^{(1)}$, we pick up the terms in equation (\ref{015}) with zeroth-order in $\eta$, thus
\ba
\label{16}
\int {\rm d} \vec r^{\prime} \;\;\left\{(1+\alpha)\rho(\partial_i^{\prime}\theta + \alpha \partial_i^{\prime}\beta)\partial_i^{\prime}\delta(\vec r^{\prime} - \vec r) -\frac{\partial {\cal G}^{(1)}}{\partial\eta}\right\}=0\,\,, \nonumber \\
\mbox{}
\ea
where $\partial_i^{\prime} =\frac{\partial}{\partial \vec r^{\prime}}$.
The linear correction term as function of $\eta$ is ${\cal G}^{(1)}=(1+\alpha)\rho\left(\partial_i\theta+\alpha\partial_i\beta\right)\partial_i\eta$.
For the quadratic correction terms in equation (\ref{015}), we have that
\be
\frac{\partial {\cal G}^{(1)}}{\partial\theta}+\frac{\partial {\cal G}^{(1)}}{\partial\beta}-\frac{\partial {\cal G}^{(2)}}{\partial\eta}=0.
\ee
And the second-order correction term is ${\cal G}^{(2)} =\frac{(1+\alpha)^2}{2}\rho\partial_i\eta\partial^i\eta$.
For the cubic correction terms in equation (\ref{015}), we can write
\ba
\frac{\partial {\cal G}^{(2)}}{\partial\theta}+\frac{\partial {\cal G}^{(2)}}{\partial\beta}-\frac{\partial {\cal G}^{(3)}}{\partial\eta}&=& 0,\nonumber\\
\frac{\partial {\cal G}^{(3)}}{\partial\eta}&=& 0,
\ea
which allows us to conclude that ${\cal G}^{(n)}=0$ for $n\geq3$.
Hence, the gauge-invariant first-order Lagrangian is written as
\ba
\label{26}
{\tilde{\cal L}}^{(0)} &=& -\rho(\dot \theta +\alpha\dot \beta)- (1+\alpha)\rho\dot\eta - {\tilde V}^{(0)},\nonumber\\
\mbox{}&=& -\rho(\dot \theta+ \dot \eta)- \alpha\rho(\dot\beta +\dot\eta) - {\tilde V}^{(0)}
\ea
where the presymplectic potential is
\ba
\label{27}
&&{\tilde V}^{(0)} =\frac{1}{2}(1-k)\rho (\partial_i \theta + \alpha\partial_i \beta) (\partial^i\theta + \alpha\partial^i\beta) \nonumber \\
&+& (1+\alpha)\rho\left(\partial_i\theta+\alpha\partial_i\beta\right)\partial_i\eta \nonumber \\
&+& \frac{(1+\alpha)^2}{2}\rho\partial_i\eta\partial^i\eta+ V(\rho),\nonumber\\
&=& \frac{1}{2}(1-k)\rho \left[\partial_i (\theta+\eta)
+  \alpha\partial_i (\beta+\eta)\right] \nonumber \\
&\times&\left[\partial^i (\theta+\eta) + \alpha\partial^i (\beta+\eta)\right]+ V(\rho)\,\,.
\ea

The Lagrangian (\ref{26}) can also be written as ${\tilde{\cal L}}^{(0)}= \rho\,\frac{\p V(\rho)}{\p \rho}$ after solving the equation of motion for $\rho$.

The zero-mode ${\tilde \nu}^{(0)}$ is the generator of infinitesimal gauge transformations $(\delta{\cal O}=\varepsilon\tilde\nu^{(0)})$. Then,
\begin{eqnarray}
\label{30}
\delta \rho(\vec r,t) &=& 0,\nonumber\\
\delta \theta(\vec r,t) &=& \varepsilon(\vec r,t)\,\,\delta(\vec r - \vec r^{\prime}),\nonumber\\
\delta \alpha(\vec r,t) &=& 0,\\
\delta \beta(\vec r,t) &=& \varepsilon(\vec r,t)\,\,\delta(\vec r - \vec r^{\prime}),\nonumber\\
\delta \eta(\vec r,t)&=& -\varepsilon(\vec r,t)\,\,\delta(\vec r - \vec r^{\prime}),\nonumber
\end{eqnarray}
where $\varepsilon(\vec r,t)$ is an infinitesimal time-dependent parameter. In fact, under the infinitesimal transformations above, the invariant Hamiltonian $({\tilde V}^{(0)})$ changes as $\delta{\tilde V}^{(0)}= 0$.
Considering the following transformations, $\tilde\theta \,=\, \theta + \eta$ and
$\tilde\beta \,=\, \beta + \eta$,
then the Lagrangian density, equation (\ref{26}), and the Hamiltonian, equation (\ref{27}), become
\ba
{\tilde{\cal L}}^{(0)} &=& -\rho(\dot{\tilde\theta}+\alpha\dot{\tilde\beta}) - {\tilde V}^{(0)}, \\
{\tilde V}^{(0)} &=& \frac{1}{2}\rho(\partial_i\tilde\theta+\alpha\partial_i\tilde\beta)(\partial^i\tilde\theta+\alpha\partial^i\tilde\beta)+ V(\rho)\,\,. \nonumber
\ea
These expressions are identical to the original expressions for the Lagrangian in equation (\ref{002}) and the Hamiltonian in equation (\ref{03}), respectively. 

Thus, at this point, it is important to point out that exist a hidden symmetry into the rotational fluid model. For the set of differential equations obtained we have no other solutions, {\it i.e.}, the model has, in fact, only one hidden symmetry. Based on the investigation done by some of us in Ref. \cite{01}, where the extra global symmetries proposed in \cite{02} are promoted to local, we can conclude that these extra global symmetries do not exist in the rotational fluid model.


\section{Final Discussions}

From the Dirac point of view, a system classified as a gauge invariant theory is one that has first-class constraints.  When a theory has second-class constraints, the gauge invariance can be recovered by converting the second-class constraints into first-class constraints.  In the literature there is a great variety of techniques, with pros and cons, to promote this kind of conversion.


In this work we are concerned not only with the gauge invariance of second-class systems but also with the obtainment of a theory physically equivalent to the original one.  We will see why we believe that symplectic embedding formalism is the most adequate technique \cite{BN}.  Besides, as demonstrated by some of us, this method has the advantage that a convenient choice of a convenient zero-mode can lead to a theory physically equivalent to the original one through the elimination of the WZ terms \cite{PD,ht}.  This means a new interpretation of the method.  However, as recently demonstrated by some of us \cite{amnowx} the choice of the zero-mode must obey some ``boundary conditions".   In other words, we can say that the physical coherence must guide us to choose the correct (or convenient) zero-mode.  But at the same time it is possible to obtain, as mentioned just above, a whole family of physically equivalent actions.

Speaking in another way, we can say that this particular mapping between the original action and the respective final gauge invariant theory can be interpreted as a kind of physical equivalence. Whenever necessary or convenient, the gauge invariance of the final actions obtained here was demonstrated via the Dirac analysis.  The characterization of the final action as a first-class system corroborates the success of this process of symplectic embedding.

Firstly in this paper we used a kind of toy model, the Proca model, to illustrate the procedure. After that, we apply the formalism to the non-linear sigma model and to the chiral Schwinger model.  In the NLSM, a hidden symmetry lying on the original phase space was disclosed, differently from other approaches \cite{BN,WN,BGB}, where the symmetry resides on the extended WZ phase space.

In the CSM, the chiral anomaly was eliminated and the gauge symmetry was recovered.
It is important to notice that this result was achieved introducing
one WZ field while other schemes in the literature thrive with the introduction of two or more WZ fields, which is the origin of the ambiguity problem.

Besides, we showed in the context of a non-Abelian model (the non-Abelian Proca model) that the symplectic embedding
formalism can be used without any restrictions concerning the noninvariant model algebra.  Other constraint conversion techniques work since the algebra was previously and necessarily taken into account.

We also have brought a gauge-invariant version for the rotational fluid model. As a consequence, we have demonstrated that the hidden symmetry found is unique.  
Although we have studied the rotational fluid model with an extra term, which introduces dissipation into the model, the results are also valid without viscosity $(k=0)$.
We have noted that, although we have dissipation, this fluid model has a hidden symmetry,
which does not belong to the other group of symmetries obtained for the irrotational fluid model \cite{01}. So, the local version of the extra global symmetries \cite{02} does not exist in the rotational fluid model, with dissipation or not. 
Furthermore, the physical meaning of the hidden symmetry can be interpreted. 


\section{Acknowledgments}

WO would like to thank CNPq (Brazilian Research Agency), for financial support.

\end{document}